\documentclass[preprint,showpacs,preprintnumbers,amsmath,amssymb,apssamp,prb,aps]{revtex4}
\usepackage{graphicx}% Include figure files
\usepackage{dcolumn}% Align table columns on decimal point
\usepackage{bm}% bold math

\newcommand{\kp}{k_{\parallel}}

\newcommand{\eb}{\varepsilon_{\rm{b}}}
\newcommand{\eps}{\varepsilon}
\newcommand{\wx}{\omega_{\textbf{\textit{k}}_{\parallel}}^{\rm x}}
\newcommand{\wph}{\omega^{\gamma}_{\textbf{\textit{k}}}}

\begin{document}

\preprint{APS/123-QED}

\title{Strong and weak coupling limits in optics of quantum well excitons}
% Force line breaks with \\
\author{C. Creatore}
\altaffiliation[Present address: ]{Dipartimento di Fisica ``A. Volta'', Universit$\grave{\rm a}$ degli Studi di Pavia, via Bassi 6, I-27100, Pavia, Italy.}
\email{creatore@fisicavolta.unipv.it}
\author{A.~L. Ivanov}%
\affiliation{School of Physics and Astronomy, Cardiff University,
Queens Buildings, CF24 3AA, Cardiff, UK}
\date{\today}
% It is always \today, today, but any date may be explicitly specified

\begin{abstract}
A transition between the strong (coherent) and weak (incoherent)
coupling limits of resonant interaction between quantum well (QW)
excitons and bulk photons is analyzed and quantified as a function
of the incoherent damping rate $\gamma_{\rm x}$ caused by
exciton-phonon and exciton-exciton scattering. For confined QW
polaritons, a second, anomalous, $\gamma_{\rm x}$-induced
dispersion branch arises and develops with increasing $\gamma_{\rm
x}$. In this case, the strong-weak coupling transition is
attributed to $\gamma_{\rm x} = \gamma_{\rm x}^{\rm tr}$ or
$\tilde{\gamma}_{\rm x}^{\rm tr}$, when the intersection of the
normal and damping-induced dispersion branches occurs either in
$\{k_{\|},\mbox{Im}[\omega],\mbox{Re}[\omega]\}$ coordinate space
(in-plane wavevector $\kp$ is real) or in
$\{\omega,\mbox{Im}[\kp],\mbox{Re}[\kp]\}$ coordinate space
(frequency $\omega$ is real), respectively. For the radiative
states of QW excitons, i.e., for radiative QW polaritons, the
transition is described as a qualitative change of the
photoluminescence spectrum at grazing angles along the QW
structure. We show that the radiative corrections to the QW
exciton states with in-plane wavevector $\kp$ approaching the
photon cone, i.e., at $\kp \rightarrow k_0\!=\!(\omega_0
\sqrt{\eb})/(\hbar c)$ ($\eb$ is the background dielectric
constant), are universally scaled by the energy parameter $\big(
\Gamma_0^2 \omega_0 \big)^{1/3}$ with $\Gamma_0$ the intrinsic
radiative width and $\omega_0$ the exciton energy at $\kp=0$,
rather than diverge. Similarly, the strong-weak coupling
transition rates $\gamma_{\rm x}^{\rm tr}$ and
$\tilde{\gamma}_{\rm x}^{\rm tr}$ are also proportional to $\big(
\Gamma_0^2 \omega_0 \big)^{1/3}$. The numerical evaluations are
given for a GaAs single quantum well with realistic parameters:
$\Gamma_0 = 45.5\,\mu$eV and $\big( \Gamma_0^2 \omega_0
\big)^{1/3}\approx 1.5$\,meV.

\end{abstract}

%Use showkeys class option if keyword

\maketitle

\section{INTRODUCTION}

Since the pioneering work by Agranovich and Dubovskii
\cite{Agranovich}, optics of quasi-two-dimensional (quasi-2D) QW
excitons became a well-established discipline. Traditionally, the
states of optically-dressed QW excitons are classified in terms of
confined quantum well polaritons (or simply QW polaritons) and
radiative quantum well polaritons (or simply the radiative states
of QW excitons). The QW polaritons are trapped and in-plane guided
by the exciton resonance, accompanied by an evanescent light field
in the growth direction (the $z$-direction), and characterized by
a single dispersion branch $\omega = \omega(\kp)$ which lies
outside the photon cone $\omega = (\hbar ck)/\sqrt{\eb}$ in the
$\{\kp,\omega\}$ plane
\cite{Nakayama,Nakayama0,Andreanib,Andreania,Jorda1,Ivchenko91}.
Although QW polaritons are invisible in standard, far-field,
experiments, they can considerably contribute to the total optical
response associated with QW excitons in a ``hidden'' way (see,
e.g., Refs.\,[\onlinecite{Ivanov97,Ivanov98,Ivanov04}]). The
radiative states of QW excitons, which refer to the radiative zone
$\kp \leq k_0$, can optically decay into bulk photon modes and
therefore are characterized by a finite radiative lifetime
\cite{Orrit,Hanamura,Andreanic,Citrin1}. The radiative states of
QW excitons have been observed and studied in many far-field
optical, both photoluminescence (PL) and scattering
(reflectivity/transmissivity), experiments (see, e.g.,
Refs.\,[\onlinecite{Feldmann,Deveaud1,Gurioli,Eccleston,Martinez,Vinattieri,Szczytko,Deveaud2,Moret}]).
In turn, confined QW polaritons can be detected by using periodic
grating \cite{Kohl} and attenuated total reflection \cite{Lagois}
techniques, as well as with near-field scanning optical microscopy
\cite{Hess} and PL imaging spectroscopy \cite{Wu,Pulizzi}.

For GaAs QWs, confined and radiative QW polaritons are usually
described in terms of $X-$, $Y-$ and $Z-$modes
\cite{Nakayama,Nakayama0,Hanamura,Andreanic,Andreanib}. The
$X-$mode states are in-plane longitudinal with polarization along
${\textbf{\textit{k}}}_{\|}$, the $Y-$mode states are in-plane
transverse with polarization normal to
${\textbf{\textit{k}}}_{\|}$, and the $Z-$mode states are
transverse with polarization along the $z$-axis. Thus the $Y-$mode
($X-$ and $Z-$mode) states couple with the $s$-polarized
($p$-polarized) light field. The $Z-$mode is absent for confined
and radiative QW polaritons associated with the heavy-hole exciton
state in a GaAs quantum well \cite{Andreanic,Andreanib}.

The main aim of the present work is to study how the resonant
coupling between quantum well excitons and bulk photons relaxes
with increasing homogeneous width $\gamma_{\rm x}$, due to
incoherent scattering of QW excitons. In particular, in order to
quantify the strong-weak coupling transition, we apply an approach
developed in Refs.\,[\onlinecite{Tait,Quattropani}] for bulk
polaritons: The transition is attributed to the intersection of
the polariton dispersion branches that results in the topological
change ``crossing'' $\leftrightarrow$ ``anti-crossing'' in the
dispersion branches, in either
$\{k_{\|},\mbox{Im}[\omega],\mbox{Re}[\omega]\}$ or
$\{\omega,\mbox{Im}[\kp],\mbox{Re}[\kp]\}$ three-dimensional (3D)
spaces (a similar paradigm has been used for
quasi-zero-dimensional polaritons in semiconductor photonic dots
\cite{Nikolaev}). The first {\it quasi-particle} case, when the
in-plane wavevector $\kp$ is real, refers to a PL experiment,
while the second {\it forced-harmonic} case, when the frequency
$\omega$ is real, can be applied to describe a scattering
(reflectivity/transmissivity) experiment. In this paper we analyze
the $\gamma_{\rm x}$-induced change of $Y-$mode QW polaritons,
which are akin to transverse bulk polaritons. Note that neither
the orthogonality between the $X-$, $Y-$ and $Z-$modes nor the
orthogonality between the radiative and confined polariton states
of the same mode is violated in the presence of the incoherent
scattering rate $\gamma_{\rm x}$, at least within the mean-field
picture we use in our study.

In contrast with bulk exciton-polaritons whose two dispersion
branches exist for any value of $\gamma_{\rm x}$ and are given by
two solutions of the bi-quadratic dispersion equation
\cite{Hopfield,Agranovich0,Tait,Quattropani}, the dispersion
equation of $Y-$mode ($s$-polarized) QW polaritons can
straightforwardly be reduced to a bi-cubic equation. As a result,
a second {\it anomalous} damping-induced dispersion branch of
confined quantum well polaritons arises and develops with
increasing $\gamma_{\rm x}$. The existence of the additional
$\gamma_{\rm x}$-induced dispersion branch is natural and allows
us to attribute the strong-weak coupling transition to the
intersection between the normal and anomalous dispersion branches.
As we clarify in this work, the intersection points refer to $\kp
= k_{\|}^{\rm tr} = k_0 - [(3 \sqrt{\eb}) / (2 \sqrt[3]{4}\,\hbar
c)]\big(\Gamma_0^2 \omega_0 \big)^{1/3} \simeq k_0$ for the
quasi-particle case and to $\omega = \omega^{\rm tr} = \omega_0 +
(\omega_0^2 \eb)/(2 c^2 M_{\rm x}) - [3/(4 \sqrt[3]{4})]\big[
(\omega_0 \eb )/(M_{\rm x} c^2) \big]^{1/3} \big(\Gamma_0^2
\omega_0 \big)^{1/3} \simeq \omega_0$ for the forced-harmonic
case, and the transition (threshold) damping rates are given by
$\gamma_{\rm x}^{\rm tr} = [(3 \sqrt{3})/(2 \sqrt[3]{4})] \big(
\Gamma_0^2 \omega_0 \big)^{1/3}$ and ${\tilde \gamma}_{\rm x}^{\rm
tr} = [(3 \sqrt{3})/2] \big[ (\omega_0 \eb )/(M_{\rm x} c^2)
\big]^{1/3} \big( \Gamma_0^2 \omega_0 \big)^{1/3}$ ($M_{\rm x}$ is
the in-plane translational mass of a QW exciton), respectively.
The above analytic expressions are surprisingly simple, taking
into account a rather non-trivial analysis of the two-fold
degeneracy of complex roots of the bi-cubic dispersion equation we
have developed in order to find $\gamma_{\rm x}^{\rm tr}$ and
${\tilde \gamma}_{\rm x}^{\rm tr}$ (similar expressions for bulk
polaritons can be obtained in a much more straightforward way, as
detailed in Refs.\,[\onlinecite{Tait,Quattropani}]).

For radiative $Y-$mode polaritons, i.e., for the radiative states
of QW excitons, which are described in terms of their radiative
width $\Gamma = \Gamma_{\rm T}(\kp)$ and (Lamb) shift $\Delta =
\Delta_{\rm T}(\kp)$, we show that the radiative width
$\Gamma_{\rm T}$ does not diverges at $\kp = k_0$ even for the
completely coherent interaction between QW excitons and bulk
photons, when $\gamma_{\rm x} = 0$. While the latter conclusion
contradicts the known result of perturbation theory
\cite{Hanamura,Andreanic,Citrin1}, i.e., that $\Gamma_{\rm T}$
diverges as $\Gamma_{\rm T} \propto 1/(k_0^2 - \kp^2)^{1/2}$ when
$\kp \rightarrow k_0$, it is consisted with numerical simulations
reported in the earlier studies \cite{Orrit,Jorda2,Popov}. We
prove the regularization of the radiative corrections at $\kp =
k_0$ and show that both $\Gamma_{\rm T}(\kp\!=\!k_0)$ and
$\Delta_{\rm T}(\kp\!=\!k_0)$ are universally scaled by $\big(
\Gamma_0^2 \omega_0 \big)^{1/3}$. There is no extra dispersion
branch, induced by damping, relevant to the radiative states of QW
excitons. However, the strong-weak coupling transition can still
be seen as a $\gamma_{\rm x}$-induced qualitative change of the
radiative corrections at $\kp \simeq k_0$ and can be observed in
photoluminescence at grazing angles along the QW structure. The
transition occurs synchronously for both radiative and confined QW
polaritons.

Thus the main results of the work are (i) regularization of the
radiative corrections to the QW exciton states at $\kp \rightarrow
k_0$, with the energy parameter $\big( \Gamma_0^2 \omega_0
\big)^{1/3}$, (ii) the $\gamma_{\rm x}$-induced anomalous
dispersion branch of QW polaritons, and (iii) a quantitative
description of the strong-weak coupling transition for resonant
interaction of bulk photons and QW excitons in the presence of
incoherent scattering. Our numerical simulations refer to a
realistic high-quality GaAs single quantum well of the width $d_z
= 25$\,nm [\onlinecite{Ivanov04}].

In Sec.\,II, a Hamiltonian relevant to resonantly interacting QW
excitons and bulk photons is outlined, and the dispersion equation
is derived by using diagram technique in order to include the
incoherent scattering rate $\gamma_{\rm x}$.

In Sec.\,III, for the case of completely coherent interaction
between QW excitons and bulk photons (the strong coupling limit,
$\gamma_{\rm x}=0$), we discuss the regularization of the
radiative corrections to the QW exciton states at $\kp \simeq k_0$
and quantify characteristic points $A$ ($\kp = k_0$) and $B$ ($\kp
= k_{\|}^{\rm (B)} > k_0$). It is shown that for point $A$ the
radiative width $\Gamma_{\rm T}$ reaches its maximum value
$\Gamma_{\rm T}^{\rm (A)} \propto \big( \Gamma_0^2 \omega_0
\big)^{1/3}$, while for the terminal point $B$ the Lamb shift
$\Delta_{\rm T}$ of the QW radiative states has a maximum value
$\Delta_{\rm T} \propto \big( \Gamma_0^2 \omega_0 \big)^{1/3}$.

In Sec.\,IV, we analyze the QW polariton states in the presence of
incoherent scattering and, in particular, prove that the second,
anomalous, QW polariton dispersion branch exists for $\gamma_{\rm
x} \geqslant \gamma_{\rm c}^{(1)} = \Gamma_0$. It is shown that
the optical brightness, i.e., the visibility, of the second
dispersion branch drastically increases when the in-plane
wavevector $\kp$ approaches $k_0$. We also quantify the threshold
damping rates, $\gamma_{\rm x}^{\rm tr}$ (quasi-particle
solutions) and $\tilde{\gamma}_{\rm x}^{\rm tr}$ (forced-harmonic
solutions), when the strong-weak coupling transition occurs, and
demonstrate that both $\gamma_{\rm x}^{\rm tr}$ and
$\tilde{\gamma}_{\rm x}^{\rm tr}$ are scaled by the energy
parameter $\big( \Gamma_0^2 \omega_0 \big)^{1/3}$.

In Sec.\,V, the transition between the strong and weak coupling
limits is analyzed for the radiative states of QW excitons. We
show that the $\gamma_{\rm x}$-induced radiative states can
persist far beyond the terminal point $B$, i.e., at $\kp >
k_{\|}^{\rm (B)}$, and that with increasing $\gamma_{\rm x}$
across $\gamma_{\rm x}^{\rm tr}$ the radiative corrections
drastically change their shape, $\Gamma_{\rm T} = \Gamma_{\rm
T}(\kp)$ and $\Delta_{\rm T} = \Delta_{\rm T}(\kp)$, in the
vicinity of $\kp = k_0$.

In Sec.\,VI, we discuss how damping-induced QW polaritons of the
anomalous dispersion branch and the strong-weak coupling
transition can be detected by using near-field optical
spectroscopy. It is also shown that the $\gamma_{\rm x}$-induced
change of the PL signal from the QW exciton radiative states at
$\kp \simeq k_0$ (photoluminescence at grazing angles) allows to
visualize the strong-weak coupling transition.

A short summary of the results is given in Sec.\,VII.

\section{MODEL}

The Hamiltonian of a system ``bulk photons -- QW excitons'', in
the presence of dipole interaction between two species, is given
by
\begin{equation}
H = H_{\gamma} + H_{\rm x} + H_{i}^{\rm I} + H_{i}^{\rm II}\,,
\label{eq:hamiltoniana1}
\end{equation}
with
\begin{eqnarray}
H_{\gamma} &=& \sum_{\textbf{\textit{k}}}
\omega^{\gamma}_{\textbf{\textit{k}}}
\alpha^{\dag}_{\textbf{\textit{k}}}
\alpha_{\textbf{\textit{k}}}\,, \ \ \ \ \ H_{\rm
x}=\sum_{\textbf{\textit{k}}_{\parallel}} \omega^{\rm
x}_{\textbf{\textit{k}}_{\parallel}}
b^{\dag}_{\textbf{\textit{k}}_{\parallel}}
b_{\textbf{\textit{k}}_{\parallel}}\,, \nonumber \\
H_{i}^{\rm I} &=& i
\sum_{\textbf{\textit{k}}_{\parallel}}\sum_{k_{z}} (\omega^{\rm
x}_{\textbf{\textit{k}}_{\parallel}})^{1/2}
C_{\textbf{\textit{k}}_{\parallel},k_{z}}
(\alpha_{\textbf{\textit{k}}_{\parallel},k_{z}}+
\alpha^{\dagger}_{-\textbf{\textit{k}}_{\parallel},-k_{z}})
(b_{-\textbf{\textit{k}}_{\parallel}}-
b^{\dagger}_{\textbf{\textit{k}}_{\parallel}})\,, \nonumber \\
H_{i}^{\rm II} &=&
 \sum_{\textbf{\textit{k}}_{\parallel}}
\sum_{k_{z},k^{'}_{z}} C_{\textbf{\textit{k}}_{\parallel},k_{z}}
C_{\textbf{\textit{k}}_{\parallel},k^{'}_{z}}
(\alpha_{\textbf{\textit{k}}_{\parallel},k_{z}} +
\alpha^{\dagger}_{-\textbf{\textit{k}}_{\parallel},-k_{z}})
(\alpha_{-\textbf{\textit{k}}_{\parallel},-k^{'}_{z}} +
\alpha^{\dagger}_{\textbf{\textit{k}}_{\parallel},k^{'}_{z}})\,,
\label{eq:H-parts}
\end{eqnarray}
where $b_{\textbf{\textit{k}}_{\parallel}}$ and
$\alpha_{\textbf{\textit{k}}}$ are the QW exciton and bulk photon
operators, respectively, $\omega^{\rm
x}_{\textbf{\textit{k}}_{\parallel}}=\omega_{0} + (\hbar^2
k_{\parallel}^{2})/(2M_{\rm x})$ and
$\omega^{\gamma}_{\textbf{\textit{k}}}=(\hbar c\textit{k})/
\sqrt{\varepsilon_{\rm b}}$ are the exciton and photon
dispersions, respectively. The coupling constant
$C_{\textbf{\textit{k}}_{\parallel},k_{z}}$ is given by
$C_{\textbf{\textit{k}}_{\parallel},k_{z}}=[R_{\rm
QW}/(2{\omega^{\gamma}_{\textbf{\textit{k}}}L})]^{1/2}$, where
$R_{\rm QW}$ is the dimensional oscillator strength of
exciton-photon interaction per QW unit area and $L$ is the
$z$-direction quantization length of the light field ($L
\rightarrow \infty$). The Hamiltonian (\ref{eq:hamiltoniana1}) is
relevant to the optics of transverse QW excitons which interact
with the in-plane TE-polarized light field ($Y$-mode). The
photon-mediated long-range exchange interaction and non-resonant
terms of QW exciton -- bulk photon coupling are included in the
description.

The quadratic Hamiltonian (\ref{eq:hamiltoniana1}) is exactly
solvable, giving rise to the quasi-2D polariton and radiative
states of QW excitons. This case deals only with coherent
interaction between the particles [$H_i^{\rm I}$ and $H_i^{\rm
II}$ terms in Eqs.\,(\ref{eq:hamiltoniana1})-(\ref{eq:H-parts})]
and therefore inherently refers to the strong coupling between QW
excitons and bulk photons. Alternatively, the (quasi-)
eigen-energies of QW excitons can be found from
Eq.\,(\ref{eq:hamiltoniana1}) by using standard diagram technique
[\onlinecite{Hanamura,Citrin1,Jorda1,Jorda2}]. The latter approach
allows us to include the particle rate $\gamma_{\rm x}$ of
incoherent scattering of QW exciton, i.e., the homogeneous
broadening. The Dyson equation for QW excitons is
\begin{equation}
G_{\textbf{\textit{k}}_{\parallel}}(\omega) =
G_{\textbf{\textit{k}}_{\parallel}}^{(0)}(\omega) +
G_{\textbf{\textit{k}}_{\parallel}}^{(0)}(\omega)
\Sigma_{\textbf{\textit{k}}_{\parallel}}(\omega)
G_{\textbf{\textit{k}}_{\parallel}}(\omega)\,,
\label{eq:Dyson-equation}
\end{equation}
where $G_{\textbf{\textit{k}}_{\parallel}}^{(0)} = 2
\omega_{\textbf{\textit{k}}_{\parallel}}^{\rm x} / [ \omega^{2} -
(\omega_{\textbf{\textit{k}}_{\parallel}}^{\rm x} - i \gamma_{\rm
x}/2)^2]$ and $G_{\textbf{\textit{k}}_{\parallel}}$ are the
propagators of optically-non-interacting and optically-dressed
excitons, respectively, and the photon-mediated self-energy is
given by
\begin{equation}
\Sigma_{\textbf{\textit{k}}_{\parallel}}(\omega) =
\frac{2\omega^{2}}{\omega_{\textbf{\textit{k}}_{
\parallel}}^{\rm x}}\sum_{k_{z}}\frac{\mid
C_{\textbf{\textit{k}}_{\parallel},k_{z}}\mid^{2}
\omega^{\gamma}_{\textbf{\textit{k}}}}{\omega^{2} -
(\omega^{\gamma}_{\textbf{\textit{k}}})^{2}}\,.
\label{eq:exciton-self-energy}
\end{equation}
Straightforward calculation of the poles of
$G_{\textbf{\textit{k}}_{\parallel}}(\omega) =
2\omega_{\textbf{\textit{k}}_{\parallel}}^{\rm x}/[\omega^{2} -
(\omega_{\textbf{\textit{k}}_{\parallel}}^{\rm x} - i\gamma_{\rm
x}/2)^2 - 2\omega_{\textbf{\textit{k}}_{\parallel}}^{\rm x}
\Sigma_{\textbf{\textit{k}}_{\parallel}}(\omega)]$, which is
obtained from
Eqs.\,(\ref{eq:Dyson-equation})-(\ref{eq:exciton-self-energy}),
yields the spectrum of optically-dressed QW excitons:
\begin{equation}
\omega^{2} - (\omega_{\textbf{\textit{k}}_{\parallel}}^{\rm x} -
i\gamma_{\rm x}/2)^2 + \frac{\varepsilon_{\rm b}R_{\rm
QW}\omega^{2}}{c^{2}\hbar^{2}
\sqrt{k_{\parallel}^{2}-k^{2}(\omega)}} = 0\,,
\label{eq:dispersion}
\end{equation}
where $k(\omega) = (\omega \sqrt{\eb})/(\hbar c)$. For
$\gamma_{\rm x}=0$, when no decoherence of exciton-photon
interaction occurs, the dispersion Eq.~(\ref{eq:dispersion}) can
also be derived by solving the Hamiltonian
(\ref{eq:hamiltoniana1}).

Both the confined and radiative states are nonperturbatively
described by the {\it same} Eq.\,(\ref{eq:dispersion}), the fact
not completely realized in literature. The QW polariton, confined
modes, which are characterized by $\mbox{\rm Im}[\omega] \leq 0$
and $\mbox{\rm Re}[\kappa] > 0$ with
$\kappa=\sqrt{k_{\|}^2-k^{2}(\omega)}$, refer to the physical
sheet of the two-fold Riemann energy plane, while the radiative
polariton modes with $\mbox{\rm Im}[\omega] \leq 0$ and $\mbox{\rm
Re}[\kappa] < 0$ are located on the unphysical sheet of the energy
plane. The latter result is a signature of the metastable states
decaying in outgoing waves [\onlinecite{Landau}]. Thus in order to
find the energy spectrum of the radiative states, one has to use
$- \kappa$ with $\mbox{\rm Re}[\kappa] > 0$ for
$\sqrt{k_{\|}^2-k^{2}(\omega)}$ in Eq.\,(\ref{eq:dispersion}).
This can also be justified by analyzing
Eq.\,(\ref{eq:Dyson-equation}) for the radiative states in terms
of the advanced, rather than retarded, Green functions. The
initial three-dimensional Hamiltonian (\ref{eq:hamiltoniana1})
maps on to a non-Hermitian two-dimensional Hamiltonian which has
the quasi-spectrum given by Eq.\,(\ref{eq:dispersion}) and
describes the localized states (confined polaritons), which are
split-off from the continuum, and the metastable states (radiative
polaritons).

The dispersion Eq.\,(\ref{eq:dispersion}) for $Y$-mode QW
polaritons was widely discussed in the last two decades for the
case $\gamma_{\rm x} = 0$
[\onlinecite{Nakayama,Nakayama0,Andreanib,Jorda1,Jorda2}]. In
contrast, the radiative states of QW excitons were mainly
considered in terms of perturbation theory
[\onlinecite{Hanamura,Andreanic,Citrin1}]. Equation
(\ref{eq:dispersion}) allows us to treat the radiative states
non-perturbatively, what is particularly important for the
vicinity of the resonant crossover between the dispersions of bulk
photons and QW excitons, i.e., when $k_{\|} \rightarrow k_0 =
k(\omega_0)$. While the main aim of the present paper is to study
how the confined and radiative states depend on the incoherent
damping rate $\gamma_{\rm x}$, in the following Section we detail
a nonperturbative analysis of the radiative states by solving
Eq.\,(\ref{eq:dispersion}) for $\gamma_{\rm x} = 0$.

\section{Non-perturbative radiative corrections to the exciton
states in quantum wells}

The radiative half-width $\Gamma/2 = \Gamma_{\rm T}/2 = -
\mbox{Im}[\omega(k_{\|})]$ and radiative (Lamb) shift $\Delta =
\Delta_{\rm T} = \mbox{Re}[\omega(k_{\|})] - \omega_0$, calculated
with Eq.\,(\ref{eq:dispersion}) for the radiative states of QW
excitons in the strong coupling limit ($\gamma_{\rm x} = 0$), are
plotted in Figs.\,1 and 2, respectively (see the solid lines). For
comparison, in Fig.\,1 the half-width $\Gamma/2$ calculated with
perturbation theory is also shown by the dashed line. In this case
the radiative width is given by
[\onlinecite{Hanamura,Andreanic,Citrin1}]
\begin{equation}
\Gamma = \Gamma_{\rm T}(\kp) =
\Gamma_{0}\,\frac{k_{0}}{\sqrt{k_{0}^{2}-k_{\parallel}^{2}}}\,,
\label{eq:gamma-per}
\end{equation}
where $\Gamma_{0} = \Gamma_{\rm T}(\kp=0) =
[\sqrt{\varepsilon_{\rm b}}/(\hbar c)]R_{\rm QW}$ is the intrinsic
optical decay rate of QW excitons at $k_{\|} = 0$. In contrast to
the perturbative approach with Eq.\,(\ref{eq:gamma-per}), the
exact radiative width of QW excitons calculated with the
Hamiltonian (\ref{eq:hamiltoniana1}) does not diverge at $k_{\|} =
k_0$ and even persists beyond the photon cone (see Fig.\,1). This
result has already been realized numerically
[\onlinecite{Orrit,Jorda2,Popov}]. Below we quantify the
characteristic points (see points $A$ and $B$ in Figs.\,1 and 2)
and clarify the origin of the regularization of $\Gamma$ in a
narrow band $k_{\|} \simeq k_0$ where the perturbative approach is
not valid anymore.

The point $A$, where a maximum value of $\Gamma = \Gamma_{\rm
T}^{\rm max}$ occurs, is specified by
\begin{equation}
k_{\|} = k_{\|}^{\rm (A)} = k_0\,, \ \ \Gamma_{\rm T}^{\rm (A)} =
\Gamma_{\rm T}^{\rm max} = {\sqrt{3} \over 2} \big( \Gamma_0^2
\omega_0 \big)^{1/3}, \ \ \mbox{and} \ \ \Delta_{\rm T}^{\rm (A)}
= {1 \over 4} \, \big( \Gamma_0^2 \omega_0 \big)^{1/3}\, .
\label{eq:A}
\end{equation}
According to Eqs.\,(\ref{eq:A}), $\Gamma_{\rm T}^{\rm max}$ is
much larger than $\Gamma_0$. The terminal point $B$, where the
radiative width $\Gamma_{\rm T}$ becomes equal to zero and the
Lamb shift reaches its maximum value $\Delta = \Delta_{\rm T}^{\rm
max}$, lies outside the photon cone, $k_{\|}^{\rm (B)} > k_0$, and
is characterized by
\begin{eqnarray}
k_{\|} &=& k_{\|}^{\rm (B)} = k_0 \left[1 + {3 \over 2} \Big( {
\Gamma_0 \over 2 \omega_0} \Big)^{2/3} \right] = k_0 + {3 \over 2
\sqrt[3]{4}}\,{\sqrt{\eb} \over \hbar c}\,\big( \Gamma_0^2
\omega_0 \big)^{1/3}\,, \nonumber\\
\Gamma_{\rm T}^{\rm (B)} &=& 0\,, \ \mbox{and} \ \ \Delta_{\rm
T}^{\rm (B)} = \Delta_{\rm T}^{\rm max} = {1 \over \sqrt[3]{4}}\,
\big( \Gamma_0^2 \omega_0 \big)^{1/3}\, . \label{eq:B}
\end{eqnarray}
In close proximity of the critical point $B$, the radiative width
$\Gamma_{\rm T}$ decreases proportionally to the square root of
$k_{\|}^{\rm (B)} - k_{\|}$ (see Fig.\,1) and is approximated by
\begin{equation}
\Gamma_{\rm T}\big(\kp \rightarrow \kp^{\rm (B)}\big) = {6
\sqrt[6]{2} \over \sqrt{5}} \, \big(\Gamma_0^2 \omega_0
\big)^{1/3} {1 \over \sqrt{k_0}} \, \sqrt{k_{\|}^{\rm (B)} -
k_{\|}} \, . \label{eq:BB}
\end{equation}
For $\gamma_{\rm x} = 0$, the case considered in this Section,
there are no roots of Eq.\,(\ref{eq:dispersion}) relevant to the
radiative modes for $k_{\|} > k_{\|}^{\rm (B)}$.

\begin{figure}[t!]
\includegraphics*[width=0.50\textwidth]{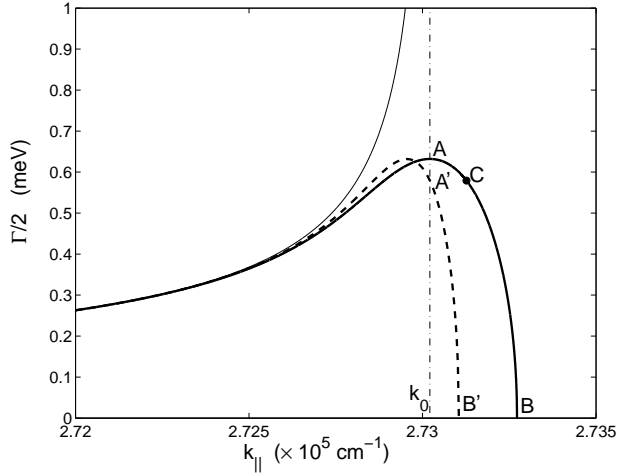}
\caption{The radiative half-width $\Gamma/2 = \Gamma_{\rm T}/2$ of
$Y$-polarized QW excitons against the in-plane wavevector $\kp$,
evaluated with the standard perturbative approach given by
Eq.\,(\ref{eq:gamma-per}) (thin solid line), the self-consistent
perturbation approach given by Eq.\,(\ref{eq:rad-rate}) (dashed line), and by the
exact diagonalization of the Hamiltonian (\ref{eq:hamiltoniana1}),
i.e., by Eq.\,(\ref{eq:dispersion}) (solid line). The
dashed-dotted vertical line indicates $k_{0}=k(\omega_{0}) =
(\sqrt{\eb} \omega_0)/(\hbar c)$. In numerical calculations we use
$R_{\rm QW}=0.025\,\mbox{eV}^2 \rm \AA$ and $\omega_0 = 1.5$\,eV,
so that $\Gamma_0 = 45.5\,\mu$eV and the intrinsic radiative
lifetime of QW excitons is given by $\tau_{\rm R} = \hbar /
\Gamma_0 = 14.5$\,ps. The critical points $A$ ($A'$), $B$ ($B'$),
and $C$ are specified in the text.}
\end{figure}

In order to understand the removal of $(k_0^2 - k_{\|}^2)^{-1/2}$
divergence which appears in the perturbative approach given by
Eq.\,(\ref{eq:gamma-per}), we examine the joint density of states
(JDS), $\rho({\bf k}_{\|},\omega)$, for the resonant optical decay
of QW excitons in the bulk photon modes:
\begin{equation}
\rho\big( \textbf{k}_{\parallel},\omega=\wx \big) \propto {1 \over
\pi} \int_{-\infty}^{+\infty}dk_{z} \, \frac{\gamma}{\gamma^{2} +
\big(\wx - \wph \big)^{2}} \, , \label{eq:JDS}
\end{equation}
where $2\gamma = \gamma_{\rm x} + \Gamma$ is the total scattering
rate of QW excitons. When $\partial\wph(\kp,k_{z})/\partial k_z
\neq 0$ at $k_z$ given by the energy conservation law $\wx(\kp) =
\wph(\kp,k_{z})$, the integrand function on the right-hand side
(r.h.s) of Eq.~(\ref{eq:JDS}) preserves its Lorentzian shape even
in terms of $k_z$. In this case $\rho(\mathbf{k}_{\parallel})$ does not depend on
$\gamma$, and the perturbative approach with
Eq.\,(\ref{eq:gamma-per}) is valid. The situation is different
when $\kp \rightarrow k_{0}$: the solution of the energy
conservation equation $\wx(\kp) = \wph(\kp=k_0,k_z)$ is $k_z=0$,
and $\partial\wph(\kp=k_0,k_z)/\partial k_z = 0$, indicating a
one-dimensional (1D) van Hove singularity in the joint density of
states. Thus $\rho(\mathbf{k}_{\parallel},\omega=\wx)$ becomes $\gamma$-dependent in
the vicinity of $\kp=k_{0}$.

\begin{figure}[t!]
\includegraphics*[width=0.50\textwidth]{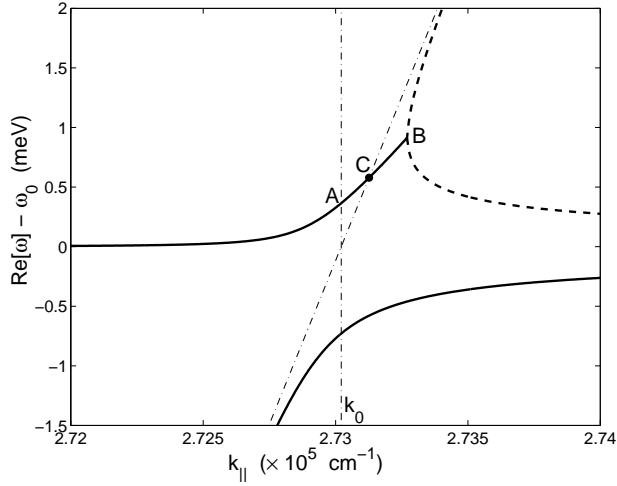}
\caption{The polariton dispersion Re$[\omega] - \omega_0$, i.e.,
the Lamb shift $\Delta = \Delta_{\rm T}$ of optically-dressed
$Y$-polarized QW excitons against the in-plane wavevector $\kp$.
The numerical calculations are done with the dispersion
Eq.\,(\ref{eq:dispersion}), the upper (lower) solid line refers to
the radiative (confined) polariton states. The dash-dotted line
shows the dispersion of bulk photons, $(\hbar c \kp/\sqrt{\eb}) -
\omega_0$. The dashed two-branch dispersion curve beyond the
terminal, bifurcation point $B$ is relevant to the radiative
states and has no physical meaning for $\gamma_{\rm x} = 0$ (see
Sec.~V). The segment $CB$ refers to the radiative states located
outside the photon cone.}
\end{figure}

By applying standard perturbation theory with the JDS determined
by Eq.\,(\ref{eq:JDS}), one receives for the optical decay of QW
excitons:
\begin{equation}
\Gamma_{\rm T}(\kp) = \frac{(\sqrt{2}k_{0}^{2} \kp
\tilde{\gamma})\Gamma_{0}} {\big[(k_{0}^{2}-\kp^{2})^{2} + 4
\tilde{\gamma}^{2} k_{0}^{2}\kp^{2}\big]^{1/2}\,
\big[[(k_{0}^{2}-\kp^{2})^{2}+ 4 \tilde{\gamma}^{2} k_{0}^{2}
\kp^{2}]^{1/2} -(k_{0}^{2}-\kp^{2})\big]^{1/2}}\, ,
\label{eq:rad-rate}
\end{equation}
where $\tilde{\gamma} = \gamma/\omega_{0}$,
$\delta=\kp/k_{0}=(\hbar c\kp)/(\sqrt{\eb} \omega_{0})$, and
$\tilde{\delta}=\sqrt{1 - \delta^{2}}$
($\tilde{\delta}=\sqrt{\delta^{2} - 1}$) for $\kp \leq k_0$ ($\kp
> k_0$). The regular perturbative solution (see the dashed line in
Fig.\,1) refers to $\kp \lesssim k_0 - \tilde{\gamma} k_0$ with
the dimensionless parameter $\tilde{\gamma} = \gamma/\omega_{0}
\sim 10^{-3}$. In this case, the JDS is given by $\rho = [(\eb
\omega_0)/(\pi c^2 \hbar^2)](k_0^2 - \kp^2)^{-1/2}$ and
Eq.\,(\ref{eq:rad-rate}) reduces to Eq.\,(\ref{eq:gamma-per}). In
contrast, for the narrow wavevector band $|\kp - k_0| \lesssim
\tilde{\gamma} k_0$ the 1D van Hove singularity strongly affects
the radiative corrections. In particular, for $\kp = k_0$
Eqs.\,(\ref{eq:JDS}) and (\ref{eq:rad-rate}) yield
\begin{equation}
\rho(\kp=k_{0})=\frac{\sqrt{\eb}}{2\pi \hbar
c}\,\frac{1}{\sqrt{\tilde{\gamma}}} \ \ \ \ \mbox{and} \ \ \ \
\Gamma_{\rm{T}}(\kp=k_{0}) =
\frac{\Gamma_{0}}{2\sqrt{\widetilde{\gamma}}}\,. \label{eq:k0}
\end{equation}
Equations (\ref{eq:k0}) clearly show that the scattering processes
relax the 1D van Hove singularity at $\kp = k_0$ and give rise to
a finite value of $\Gamma_{\rm{T}}(\kp=k_{0}) \propto
1/\gamma^{1/2}$.

For the completely coherent interaction of QW excitons with bulk
photons, when $\gamma_{\rm x} = 0$, Eq.\,(\ref{eq:rad-rate}) can
also be interpreted in terms of ``self-consistent'' perturbation
theory. In this case, $\tilde{\gamma} = \Gamma_{\rm T}/(2
\omega_0)$ and for the band $|\kp - k_0| \lesssim \tilde{\gamma}
k_0$ the radiative width $\Gamma_{\rm T} = \Gamma_{\rm T}(\kp)$
should be found as a solution of Eq.\,(\ref{eq:rad-rate}) with the
right-hand side (r.h.s.) explicitly dependent on $\Gamma_{\rm T}$.
This solution is well approximated by using a real root of the
cubic equation, $x^2(x + \eps) = \Gamma_0^2/(16 \omega_0^2)$, with
$x = \sqrt{\eps^2 + \Gamma_{\rm T}^2/(4 \omega_0^2)}$ and $\eps =
(\kp - k_0)/k_0$ ($|\eps| \lesssim \tilde{\gamma}$). The radiative
width $\Gamma_{\rm T} = \Gamma_{\rm T}(\kp)$ numerically evaluated
with the self-consistent perturbation theory is shown in Fig.\,1
by the dashed line. The self-consistent perturbation theory
reproduces qualitatively the exact solution of
Eq.\,(\ref{eq:dispersion}) for the radiative states (solid line in
Fig.\,1). In particular, for $\kp = k_0$ the self-consistent
perturbative Eqs.\,(\ref{eq:k0}) with $\tilde{\gamma} =
\Gamma_{\rm T}/(2 \omega_0)$ yield $\Gamma_{\rm T}(\kp=k_0) =
(1/\sqrt[3]{2})\,\big( \Gamma_0^2 \omega_0 \big)^{1/3}$, a value
by only 8\% smaller than the exact one, $\Gamma_{\rm T}^{\rm
(A)}$, given by Eqs.\,(\ref{eq:A}) (see points $A$ and $A'$ in
Fig.\,1).

It is the scattering-induced relaxation of energy conservation in
the resonant conversion ``QW exciton $\leftrightarrow$ bulk
photon'', $\omega_{\textbf{\textit{k}}_{\parallel}}^{\rm
x}~(\simeq \omega_0) = \hbar (c/\sqrt{\eb})\sqrt{\kp^{2} +
k_{z}^{2}}$, that is responsible for the appearance of the
radiative states beyond the photon cone, i.e., beyond the point
$C$ in Figs.\,1 and 2. Furthermore, even for $\gamma_{\rm x}=0$
the coherent optical decay itself relaxes energy conservation,
putting the quasi-eigenenergies of the radiative states in the
complex plane (of course, the energy is strictly conserved in the
incoming and outgoing channels of the scattering process
``incoming bulk photon $\rightarrow$ QW exciton $\rightarrow$
outgoing bulk photon''). A straightforward analysis of
Eq.\,(\ref{eq:JDS}) shows the existence of the JDS relevant to the
resonant coupling of QW excitons and bulk photons even beyond the
photon cone, if $\gamma$ is nonzero, and qualitatively justifies
the asymptotic $\Gamma_{\rm T} \propto \sqrt{k_{\|}^{\rm (B)} -
\kp}$ [see Eq.\,(\ref{eq:BB})] which is valid for $k_0 < \kp \leq
k_{\|}^{\rm (B)}$.

For the strong coupling limit of QW excitons and bulk photons
($\gamma_{\rm x} = 0$), the width $\Gamma_{\rm T}$ and Lamb shift
$\Delta_{\rm T}$ of the radiative states at $\kp \simeq k_0$ are
uniquely scaled by the control parameter $\big( \Gamma_0^2
\omega_0 \big)^{1/3}$ [see Eqs.\,(\ref{eq:A})-(\ref{eq:B})].
Furthermore, the maximum radiative width $\Gamma_{\rm T}^{\rm max}
= \Gamma_{\rm T}^{\rm (A)}$ cannot screen completely the maximum
radiative blue shift $\Delta_{\rm T}^{\rm max} = \Delta_{\rm
T}^{\rm (B)}$, because the shift -- half-width ratio $2\Delta_{\rm
T}^{\rm max}/\Gamma_{\rm T}^{\rm max} = (2\sqrt[3]{2})/\sqrt{3}
\simeq 1.5$, according to Eqs.\,(\ref{eq:A})-(\ref{eq:B}). The
radiative corrections can be seen experimentally for high-quality
GaAs quantum wells even with a relatively small oscillator
strength of QW excitons. For example, for the parameters used in
our evaluations we estimate $\big( \Gamma_0^2 \omega_0 \big)^{1/3}
\simeq 1.46$\,meV, $\Gamma_{\rm T}^{\rm max}/2 \simeq 0.63$\,meV,
and $\Delta_{\rm T}^{\rm max} \simeq 0.92$\,meV.

\section{Strong-weak coupling transition for quantum well polaritons}

In this Section, in order to study how the QW polariton effect
relaxes with increasing damping, we analyze the dispersion
Eq.\,(\ref{eq:dispersion}) with a nonzero excitonic damping
$\gamma_{\rm x}$. Following the terminology developed by
Tait\,\cite{Tait} for bulk polaritons, two cases are
distinguished: the quasi-particle solution $\omega =
\omega(k_{\|})$ (wavevector $k_{\|}$ is real) and the
forced-harmonic solution $k_{\|} = k_{\|}(\omega)$ (frequency
$\omega$ is real).

\subsection{Quasi-particle solutions for quantum well polaritons}

For QW polaritons, the true solutions $\omega =\omega(\kp)$ of
Eq.\,(\ref{eq:dispersion}) have to satisfy the following
conditions: $\mbox{Re}[\kappa] \equiv \mbox{Re}\big[\sqrt{\kp^{2}
- k^{2}(\omega)}\big] \geq 0$ and $\mbox{Im}[\omega] \leq 0$. The
first criterion ensures that the light field associated with QW
polaritons has an evanescent envelope in the $z$-direction,
$E(z)=E(0)\exp(-\kappa|z|)$, while the second one stems from the
casuality principle. For small $\gamma_{\rm x}$, there is only one
dispersion branch which is relevant to QW polaritons, $\omega =
\omega_{1}(\kp)$ (see the r.h.s. solid line $n$ in Fig.\,3), as
detailed, e.g., in
Refs.~[\onlinecite{Orrit,Nakayama,Andreania,Andreanib}] for
$\gamma_{\rm x} = 0$. However, for $\gamma_{\rm x} = \gamma_{\rm
c}^{(1)}$ a new, second dispersion branch of QW polaritons,
$\omega=\omega_{2}(k_{\|})$, emerges and develops with increasing
$\gamma_{\rm x} \geq \gamma_{\rm c}^{(1)}$ (see the left-hand side
lines $a$ in Fig.\,3). The critical rate $\gamma_{\rm c}^{(1)}$ of
incoherent scattering is given by
\begin{equation}
\gamma_{\rm x} = \gamma_{\rm c}^{(1)} = R_{\rm QW} \,
\frac{\sqrt{\varepsilon_{\rm b}}}{\hbar c} = \Gamma_0 \,,
\label{eq:cr1}
\end{equation}
i.e., is exactly equal to the intrinsic radiative width of QW
excitons with $\kp=0$. The second dispersion branch has two
symmetric terminal points $A_{\rm f}'$ and $A_{\rm f}$ at $\kp =
\pm \kp^{\rm f}$ (the point $A_{\rm f}'$ is not shown in Fig.\,3),
where $\kp^{\rm f}$ is given by
\begin{equation}
\kp^{\rm f} = \kp^{\rm f}(\gamma_{\rm x} \geq \gamma_{\rm
c}^{(1)}\!=\!\Gamma_0) = \omega_0 \frac{\sqrt{\varepsilon_{\rm
b}}}{\hbar c} \sqrt{1 - {\varepsilon_{\rm b} R_{\rm QW}^2 \over
\gamma_{\rm x}^2 \hbar^2 c^2}} = k_0 \sqrt{ 1 - \left({\Gamma_0
\over \gamma_{\rm x}} \right)^2}\,, \label{eq:cr2}
\end{equation}
The terminal points $A_{\rm f}'$ and $A_{\rm f}$ are characterized
by Re$[\omega_2(k_{\|}\!=\!\pm k_{\|}^{\rm f})] =
\sqrt{\omega_{\textbf{\textit{k}}_{\parallel}}^{\rm x} -
\gamma_{\rm x}^2} \simeq \omega_0$, Im$[\omega_2(k_{\|}\!=\!\pm
k_{\|}^{\rm f})] = 0$ and $\textrm{Re}[\kappa(\kp\!=\!\pm \kp^{\rm
f})]=0$. In addition, the middle point $A_{\rm i}$ of the segment
$A_{\rm f}'A_{\rm f}$ is specified by $\kp= \kp^{\rm i} = 0$ and
Re$[\omega_2(\kp\!=\!0)] = \sqrt{\omega_0^2 - \Gamma_0^2} \simeq
\omega_0$ (see Fig.\,3).

\begin{figure}[t!]
\includegraphics*[width=0.50\textwidth]{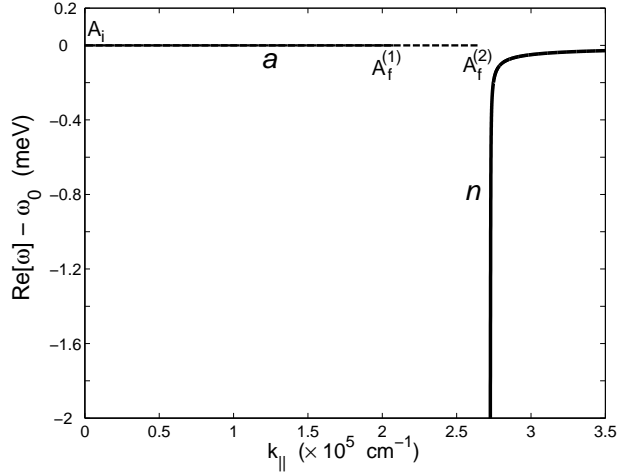}
\caption{The quasi-particle dispersion branches of QW polaritons,
$\mbox{Re}[\omega] = \mbox{Re}[\omega(k_{\|})]$, evaluated with
Eq.\,(\ref{eq:dispersion}). The solid curve $n$ shows the normal
QW polariton dispersion branch calculated for $\gamma_{\rm x} =
0$. The solid (dashed) curve $a$ refers to the anomalous,
$\gamma_{\rm x}$-induced QW polariton dispersion branch calculated
for $\gamma_{\rm x} = 70\,\mu$eV ($\gamma_{\rm x} = 200\,\mu$eV).
The critical damping $\gamma_{\rm c}^{(1)} = \Gamma_0 =
45.5\,\mu$eV.}
\end{figure}

At marginal points $A_{\rm f}'$ and $A_{\rm f}$ (see Fig.\,3),
which are characterized by Im$[\omega_2] = 0$, the anomalous,
damping-induced dispersion branch $\omega = \omega_2(k_{\|})$
appears from and leaves for the unphysical part of 3D space
$\{k_{\|},\mbox{Im}[\omega],\mbox{Re}[\omega]\}$. The dispersion
Eq.\,(\ref{eq:dispersion}) can easily be transformed to a cubic
equation, generally with complex coefficients, for $\omega^2 =
\omega^2(\kp)$. For $\gamma_{\rm x} = 0$, when the coefficients
are real, one of the roots of this equation is characterized by
Im$[\omega^2] = 0$ and corresponds to QW polaritons, while second
and third roots are complex conjugated. Among the latter two
solutions, one of the roots satisfies the selection criteria for
radiative polaritons, as discussed in Sec.\,V, giving rise to the
radiative state, while another one is unphysical. For $\gamma_{\rm
x} \geq \gamma_{\rm c}^{(1)} = \Gamma_0$, the unphysical branch
evolves in $\{k_{\|},\mbox{Im}[\omega],\mbox{Re}[\omega]\}$ space
in such a way that it has a segment where the criteria for QW
polaritons are met. Similarly to the radiative states which can
exist beyond the photon cone, the appearance of the second
dispersion branch of QW polaritons within the photon cone is due
to relaxation of the energy conservation $\delta$-function by the
damping rate $\gamma_{\rm x}$. Because $\mbox{Re}[\omega_2] \simeq
\omega_0$, the anomalous dispersion branch can be interpreted in
terms of a new, damping-induced optical decay channel of the
exciton states that opens up and develops with increasing
$\gamma_{\rm x} \geq \gamma_{\rm c}^{(1)}$. In this case, a QW
exciton with momentum $k_{\|} < k_0$, i.e., within the photon
cone, can directly emit an interface photon of the evanescent
light field. The damping-induced dispersion branches are known in
plasma physics [\onlinecite{Shivarova}] and in physics of the
surface electromagnetic waves [\onlinecite{Halevi1,Halevi2}].

\begin{figure*}[t!]
\resizebox{\hsize}{!}{
\begin{tabular}{c c c}
\includegraphics*[width=0.26\textwidth]{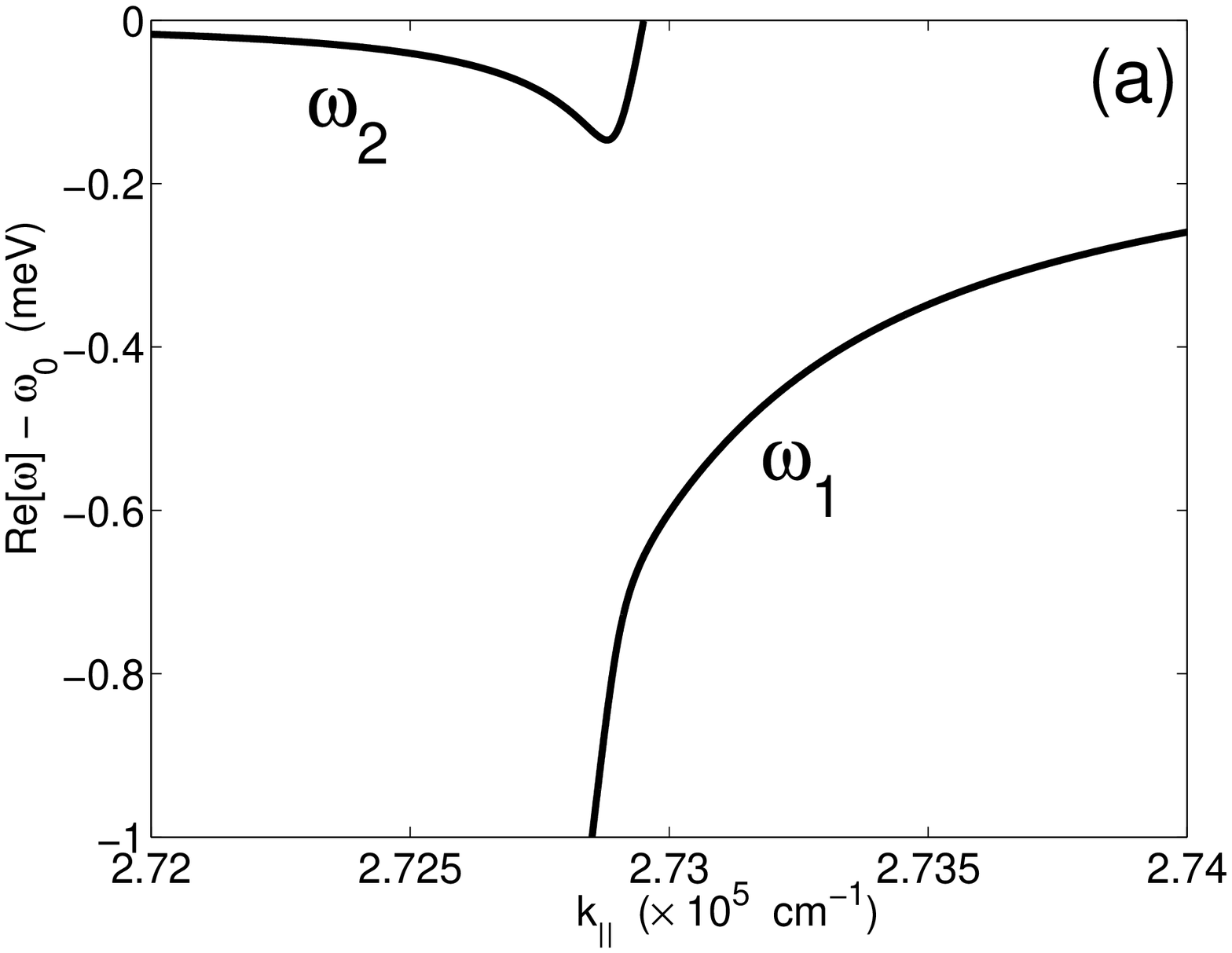}&
\includegraphics*[width=0.26\textwidth]{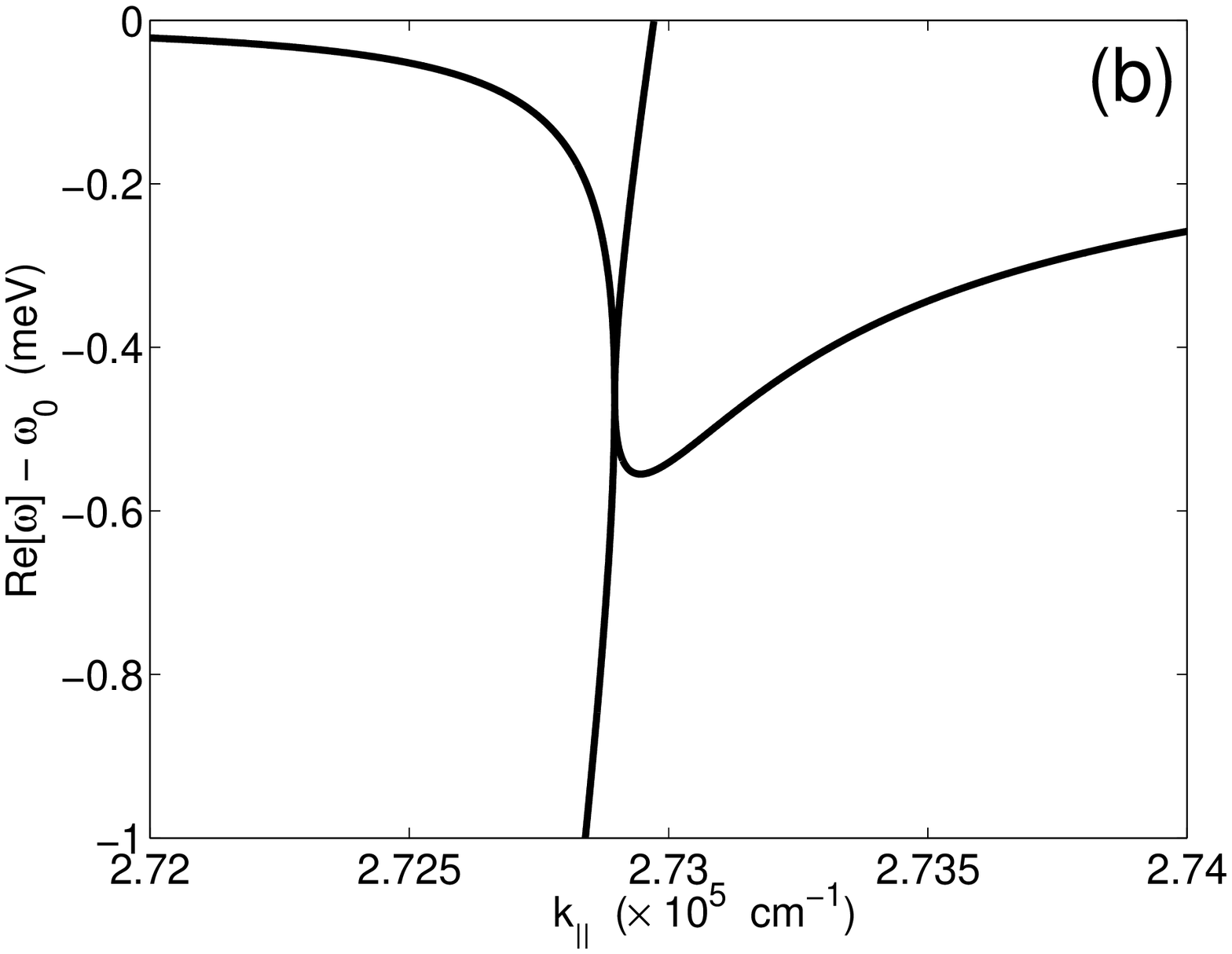}&
\includegraphics*[width=0.26\textwidth]{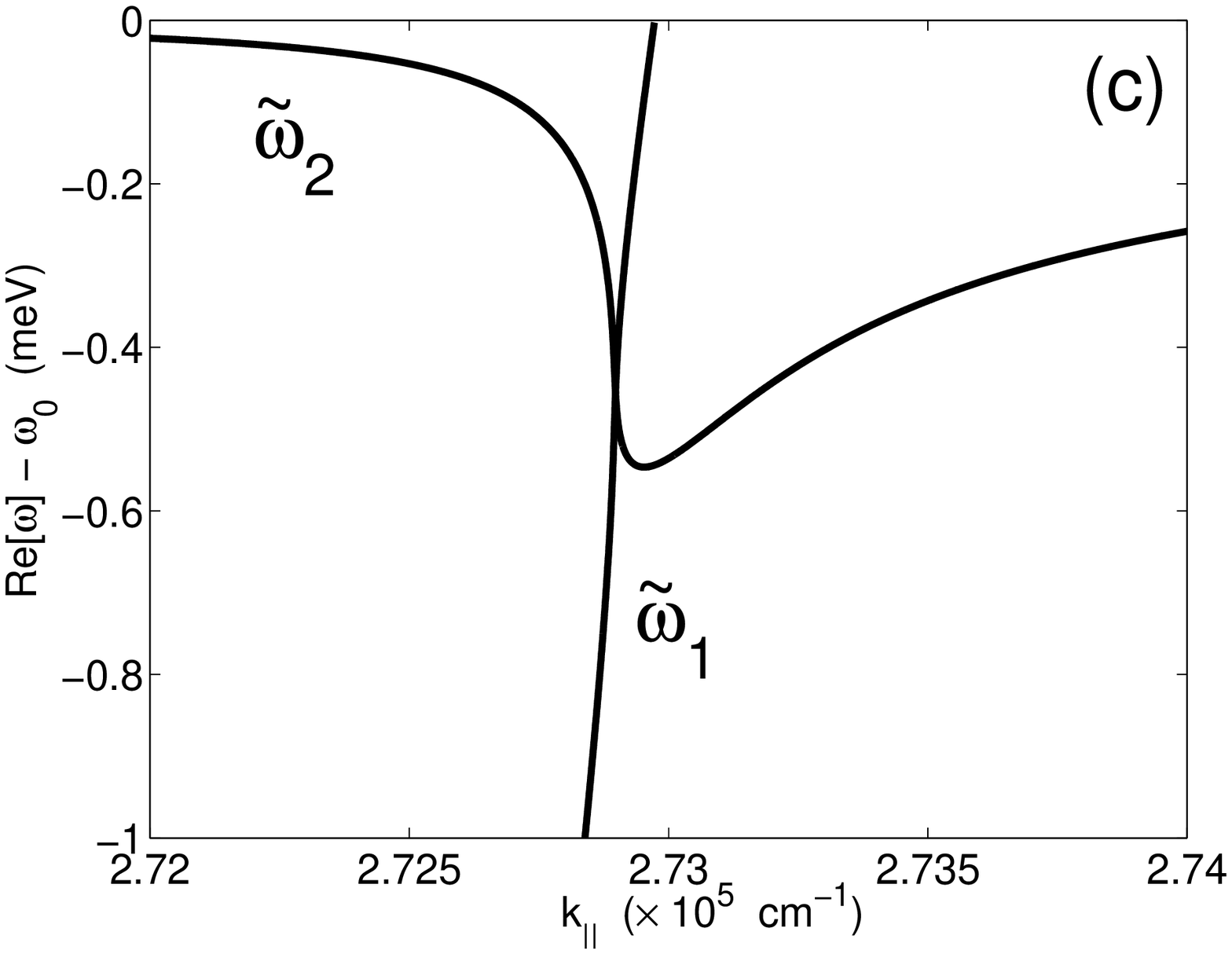}\\
\end{tabular} }\\
\resizebox{\hsize}{!}{
\begin{tabular}{c c c}
\includegraphics*[width=0.26\textwidth]{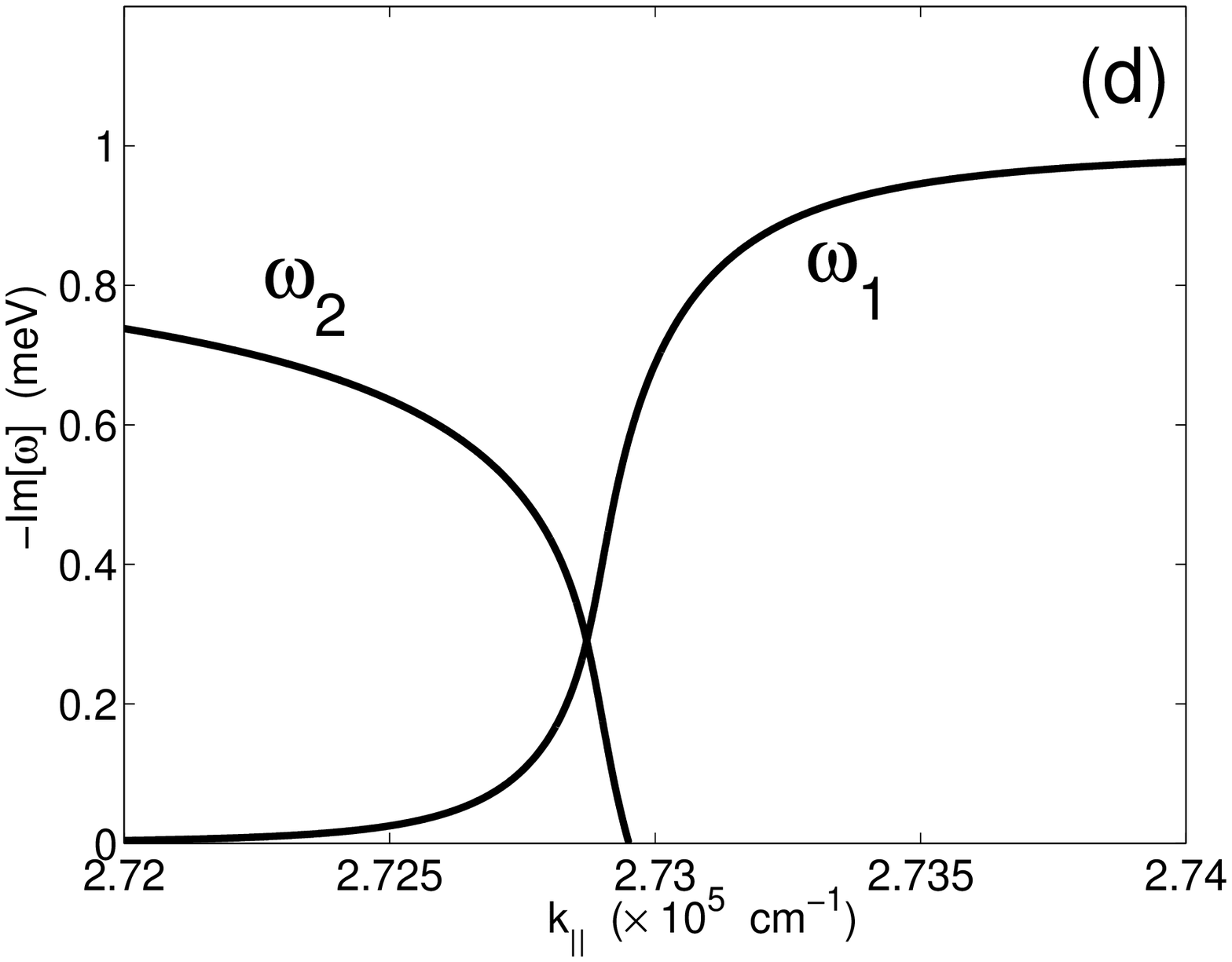}&
\includegraphics*[width=0.26\textwidth]{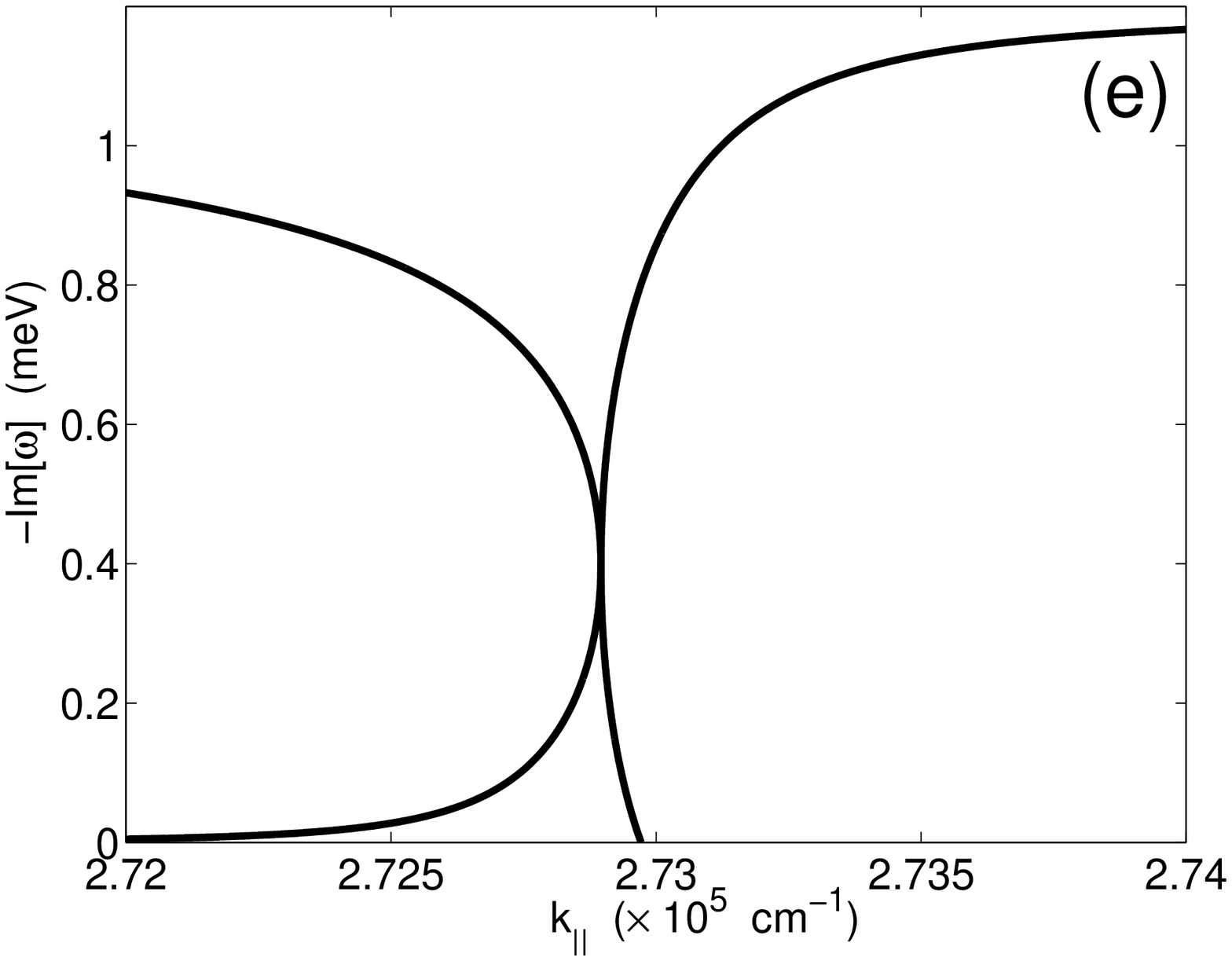}&
\includegraphics*[width=0.26\textwidth]{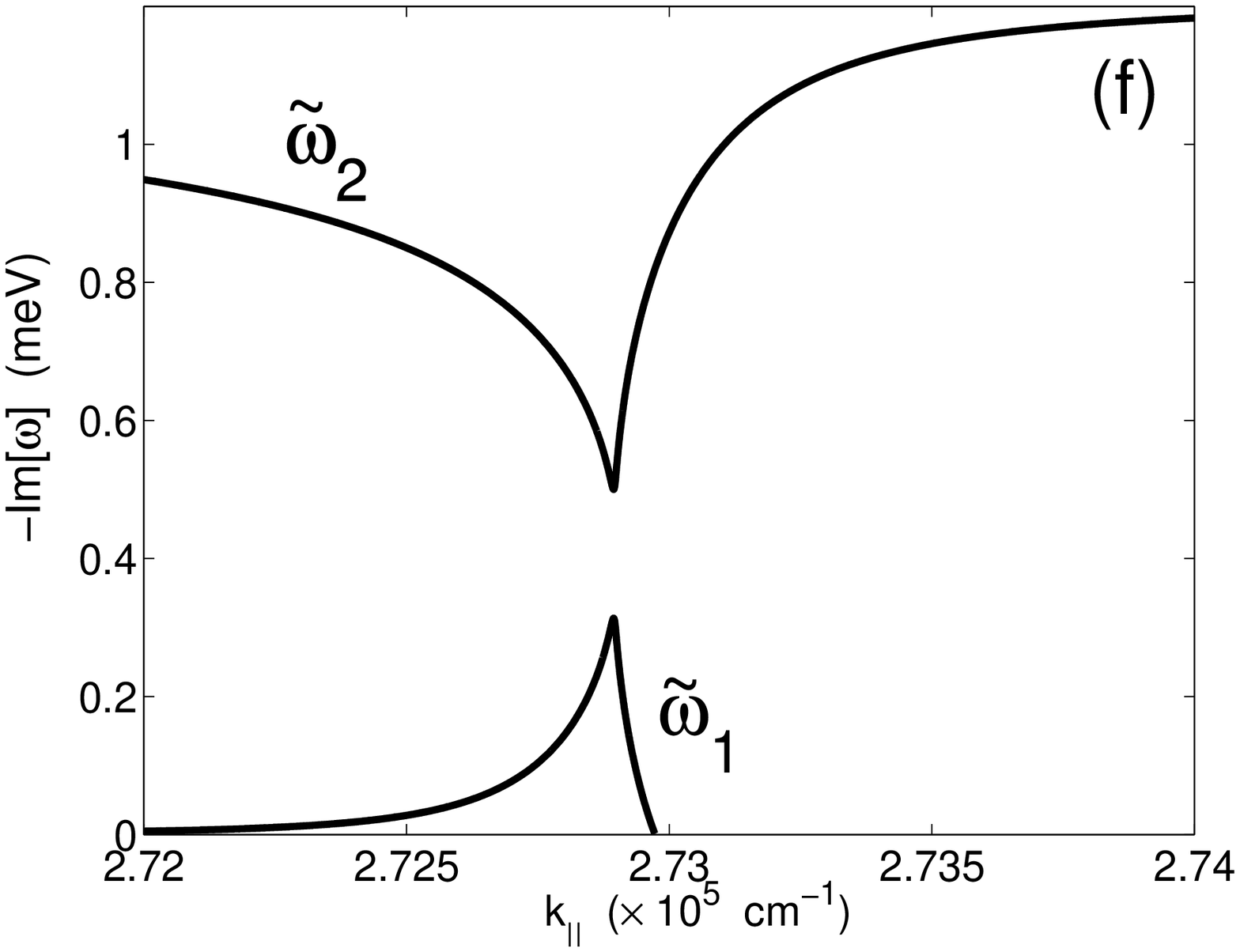}\\
\end{tabular}}
\resizebox{\hsize}{!}{
\begin{tabular}{c c c}
\includegraphics*[width=0.26\textwidth]{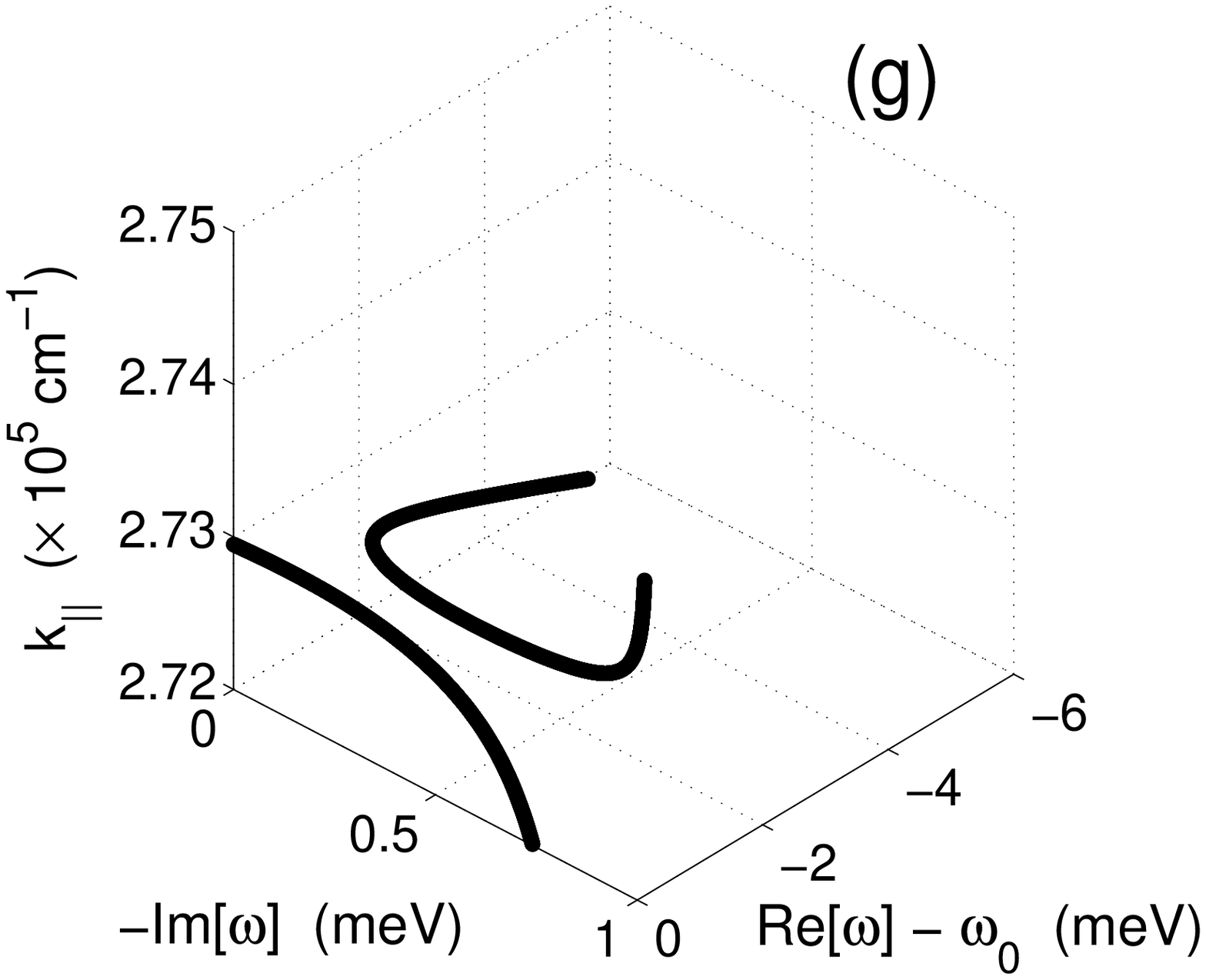}&
\includegraphics*[width=0.26\textwidth]{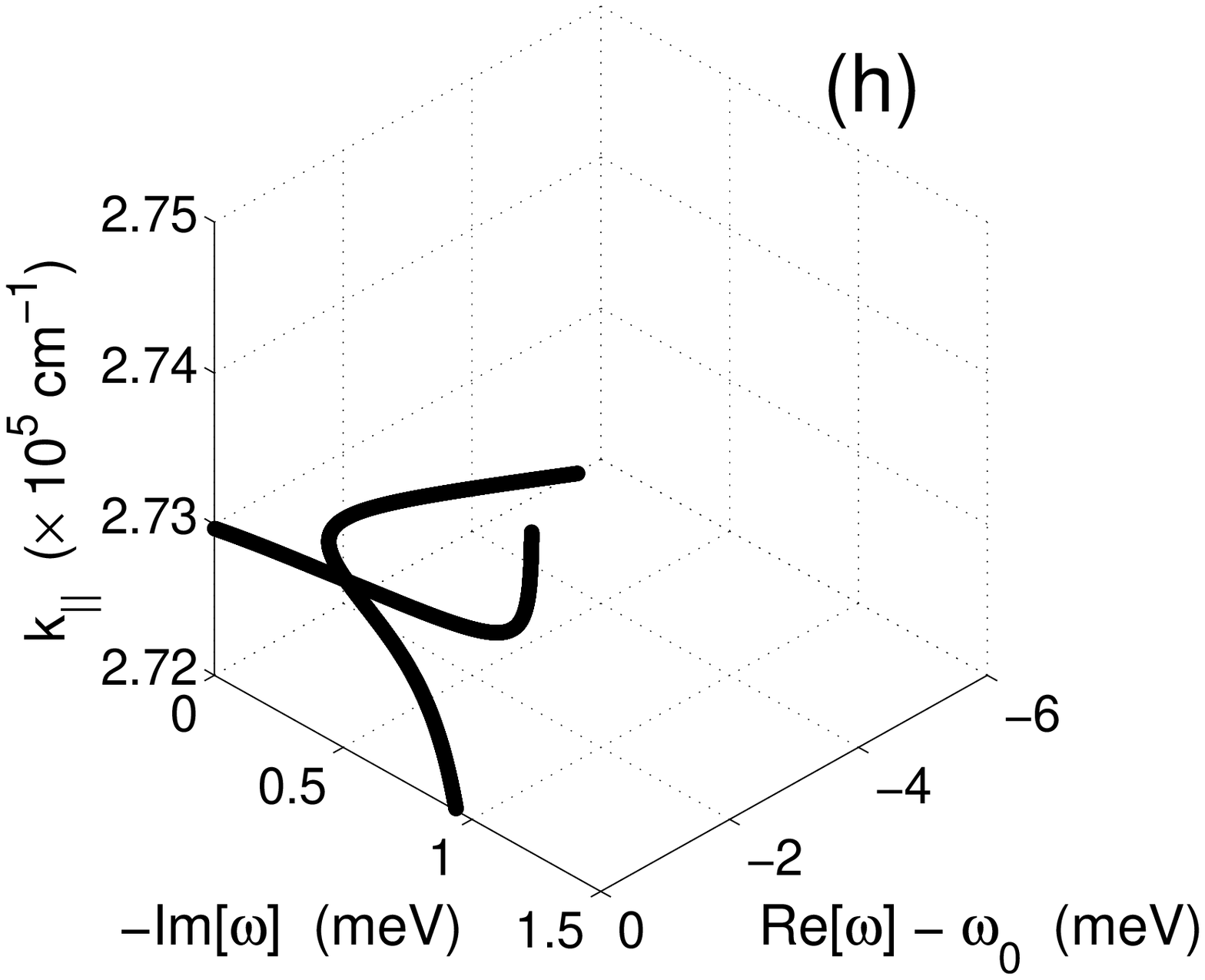}&
\includegraphics*[width=0.26\textwidth]{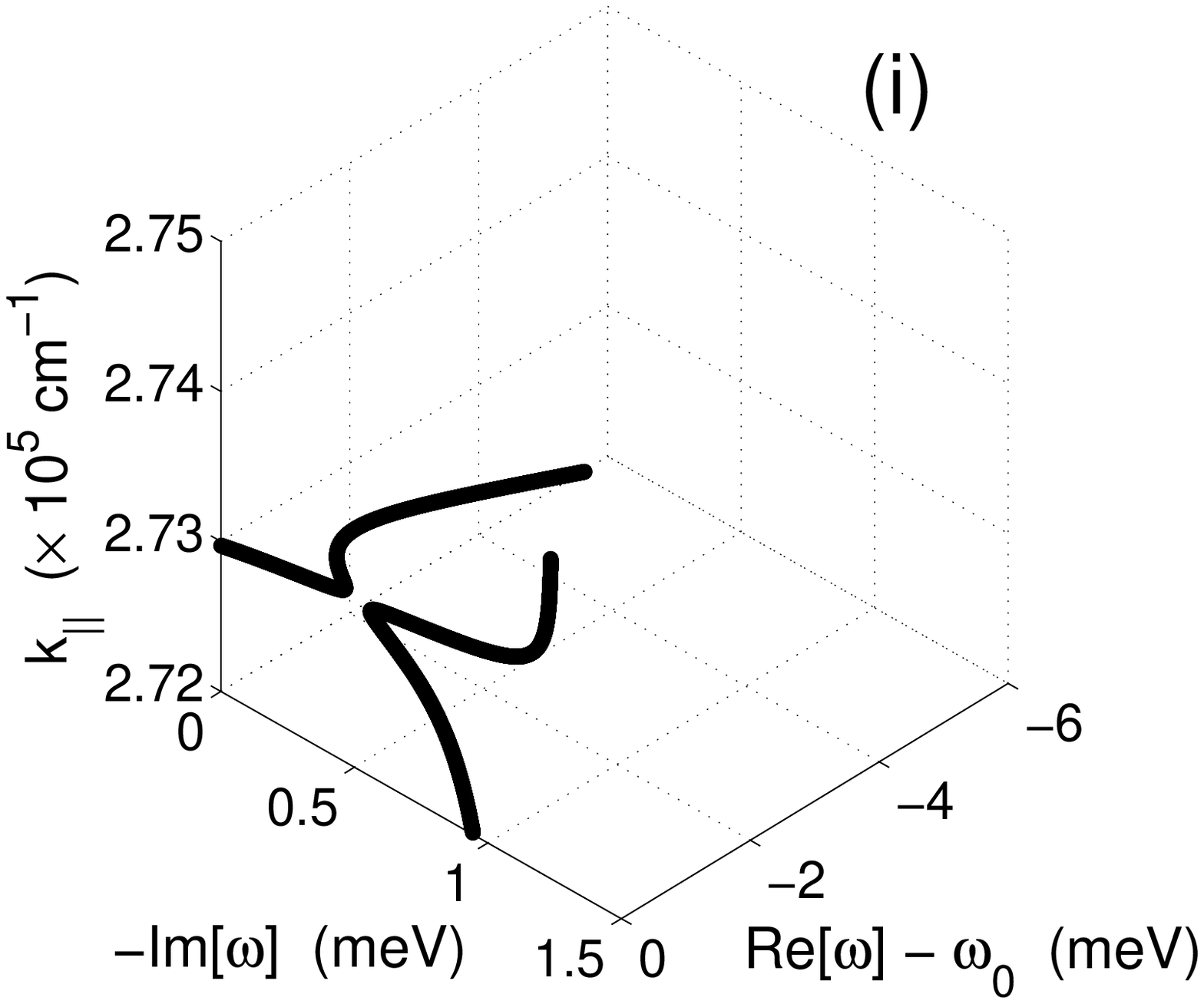}\\
\end{tabular}}
\caption{The damping-induced transition between the strong and
weak coupling limits for quasi-particle QW polaritons [$\omega =
\omega(k_{\|})$ is a solution of the dispersion
Eq.\,(\ref{eq:dispersion}) parametrically dependent upon real
$k_{\|}$]. For $R_{\rm QW} = 0.025\,\mbox{eV}^2 \rm \AA$, the
transition damping rate is given by $\gamma_{\rm x}^{\rm tr} =
\gamma_{\rm c}^{(2)} \simeq 2.39$\,meV, according to
Eqs.\,(\ref{eq:trans1}). (a), (d), (g): The strong coupling
regime, $\gamma_{\rm x} = 2.20\,\mbox{meV} < \gamma_{\rm
c}^{(2)}$. (b), (e), (h): The transition point, $\gamma_{\rm x} =
\gamma_{\rm x}^{\rm tr} = \gamma_{\rm c}^{(2)}$. (c), (f), (i):
The weak coupling regime, $\gamma_{\rm x} = 2.44\,\mbox{meV}
> \gamma_{\rm c}^{(2)}$.}
\end{figure*}

With increasing $\gamma_{\rm x} \geq \gamma_{\rm c}^{(1)}$, the
two dispersion curves, $\omega_1 = \omega_1(k_{\|})$ and $\omega_2
= \omega_2(k_{\|})$, move in 3D space
$\{k_{\|},\mbox{Im}[\omega],\mbox{Re}[\omega]\}$ towards each
other (see Fig.\,4). The intersection of the dispersion curves,
which occurs for $\gamma_{\rm x} = \gamma_{\rm x}^{\rm tr} =
\gamma_{\rm c}^{(2)}$, is interpreted as a transition between the
strong ($\gamma_{\rm x} \leq \gamma_{\rm c}^{\rm tr}$) and weak
($\gamma_{\rm x} \geq \gamma_{\rm c}^{\rm tr}$) coupling limits of
QW exciton -- photon interaction. According to the dispersion
Eq.~(\ref{eq:dispersion}), there is only one intersection point,
which is given by the conditions
$\mbox{Re}[\omega_1(k_{\|},\gamma_{\rm x}^{\rm tr})] =
\mbox{Re}[\omega_2(k_{\|},\gamma_{\rm x}^{\rm tr})]$ and
$\mbox{Im}[\omega_1(k_{\|},\gamma_{\rm x}^{\rm tr})] =
\mbox{Im}[\omega_2(k_{\|},\gamma_{\rm x}^{\rm tr})]$ [see
Figs.\,4(b), 4(e) and 4(h)]. These conditions define the
transition parameters:
\begin{eqnarray}
\gamma_{\rm x}^{\rm tr} &=& \gamma_{\rm c}^{(2)} = \frac{ 3
\sqrt{3}}{2}\,\left( \frac{\omega_0 R_{\rm QW}^2 \varepsilon_{\rm
b}}{4 c^2 \hbar^2}\right)^{1/3} \! \! = {3 \sqrt{3} \over 2
\sqrt[3]{4} } \, \big( \Gamma_0^2 \omega_0 \big)^{1/3} \simeq
1.64\,\big( \Gamma_0^2 \omega_0 \big)^{1/3}\,,
\nonumber \\
k_{\|}^{\rm tr} &=& k_0 \left[1 - {1 \over \sqrt{3}} \,
{\gamma_{\rm c}^{(2)} \over \omega_0} \right] = k_0 - {3 \over 2
\sqrt[3]{4}}\,{\sqrt{\eb} \over \hbar c}\,\big( \Gamma_0^2
\omega_0 \big)^{1/3}\,. \label{eq:trans1}
\end{eqnarray}
The transition point is characterized by the complex polariton
frequency $\omega(k_{\|}^{\rm tr}) = \omega_1(k_{\|}^{\rm tr}) =
\omega_2(k_{\|}^{\rm tr})$ given by
\begin{eqnarray}
\Delta_{\rm T}^{\rm tr} &=& \mbox{Re}[\omega(k_{\|}^{\rm tr})] -
\omega_0 = - {1 \over \sqrt[3]{4}}\,\big( \Gamma_0^2 \omega_0
\big)^{1/3}\,,
\nonumber \\
\Gamma_{\rm T}^{\rm tr} &=& -2\,\mbox{Im}[\omega(k_{\|}^{\rm tr})]
= {2 \over 3}\,\gamma_{\rm c}^{(2)} = {\sqrt{3} \over \sqrt[3]{4}
} \, \big( \Gamma_0^2 \omega_0 \big)^{1/3}\,. \label{eq:trans2}
\end{eqnarray}
For the parameters used in our numerical evaluations, $\gamma_{\rm
x}^{\rm tr} = \gamma_{\rm c}^{(2)} \simeq 2.39$\,meV, $\Delta_{\rm
T}^{\rm tr} \simeq - 0.92$\,meV, and $\Gamma_{\rm T}^{\rm tr}
\simeq 1.59$\,meV.

At the intersection point between two dispersion curves, when
$\gamma_{\rm x} = \gamma_{\rm x}^{\rm tr}$, the interconnection
between the dispersion branches changes from ``anti-crossing''
(strong coupling regime) to ``crossing'' (weak coupling regime).
As a result, topologically new dispersion branches,
$\tilde{\omega}_1(k_{\|})$ and $\tilde{\omega}_2(k_{\|})$, arise
for $\gamma_{\rm x} > \gamma_{\rm x}^{\rm tr}$ (see Fig.\,4). For
example, for $\gamma_{\rm x} = 0$ the only QW polariton dispersion
branch $\omega_1 = \omega_1(k_{\|})$ can be interpreted in terms
of photon-like [$k_{\|} \lesssim k_0 = k_0$] and exciton-like
($k_{\|} \gtrsim k_0$) parts [``anti-crossing'' of the exciton and
photon dispersions, see Fig.\,4\,(a)]. In contrast, for
$\gamma_{\rm x} \gtrsim \gamma_{\rm x}^{\rm tr}$, i.e., after the
transition, the new dispersion branch $\tilde{\omega}_{1}(\kp)$,
which starts at $k_{\|} = 0$ with $\tilde{\omega}_1 = 0$ and
terminates at $k_{\|} = k_{\|}^{\rm f}(\gamma_{\rm x}) \simeq k_0$
[for $\gamma_{\rm x} \gg \gamma_{\rm c}^{(2)} = \gamma_{\rm
x}^{\rm tr}$, Eq.\,(\ref{eq:cr2}) yields $k_{\|}^{\rm f}
\rightarrow k_0$] with $\tilde{\omega}_1 \simeq \omega_0$, can be
visualized as a purely photon-like branch [``crossing'' of the
exciton and photon dispersions, see Fig.\,4\,(c)]. In a similar
way, the entire $\tilde{\omega}_2$-branch can be interpreted in
terms of the exciton dispersion.

\subsection{Forced-harmonic solutions for quantum well polaritons}

For quasi-2D QW polaritons, the forced-harmonic solutions $\kp =
\kp(\omega)$ of Eq.\,(\ref{eq:dispersion}) should satisfy the
following conditions: Re$[\kappa] \geq 0$ and Im$[\kp] \geq 0$.
Comparing with the quasi-particle solutions, an additional
dimensionless parameter $\nu_{\rm x} = (\omega_0 \eb)/(M_{\rm x}
c^2)$ is relevant to this class of solutions. The parameter
$\nu_{\rm x}$ explicitly depends on the translational mass of QW
excitons, so that $\nu_{\rm x} \rightarrow 0$ for $M_{\rm x}
\rightarrow \infty$. For the control parameters relevant to GaAs
QWs, one estimates $\nu_{\rm x} \simeq (1.2 - 1.3) \times
10^{-4}$.

Similarly to the quasi-particle solutions, a new dispersion branch
$\kp^{(2)}=\kp^{(2)}(\omega)$ emerges with increasing $\gamma_{\rm
x} \geq \tilde{\gamma}_{\rm c}^{(1)}$. In this case, however,
$\tilde{\gamma}_{\rm c}^{(1)} = 0$, and the anomalous dispersion
branch starts to develop from $\{\kp^{\rm i}\!=\!0,\omega^{\rm
i}\!=\!0\}$ point. For a given incoherent scattering rate
$\gamma_{\rm x} > 0$, the terminal point $A_{\rm f}$, where the
$\gamma_{\rm x}$-induced branch leaves for the unphysical part of
3D space $\{\omega,\mbox{Im}[k_{\|}],\mbox{Re}[k_{\|}]\}$ (see
Figs.\,5 and 6), is characterized by the frequency
\begin{equation}
\omega = \omega^{\rm f}(\gamma_{\rm x}) = \frac{\omega_0}{\big[ 1
- \nu_{\rm x} + \nu_{\rm x} (\Gamma_0 /\gamma_{\rm x})^2
\big]^{1/2}}\,. \label{eq:hf1}
\end{equation}
Equation (\ref{eq:hf1}), which is valid for $\nu_{\rm x} \ll 1$,
indeed shows that $\tilde{\gamma}_{\rm c}^{(1)} = 0$: $\omega^{\rm
f} \rightarrow 0$ for $\gamma_{\rm x} \rightarrow 0$. For
$\gamma_{\rm x} \gtrsim \Gamma_0$, when frequency $\omega^{\rm f}$
approaches $\omega_0$ (see Fig.\,5), Eq.\,(\ref{eq:hf1}) reduces
to
\begin{equation}
\omega = \omega^{\rm f}(\gamma_{\rm x}\!\gtrsim\,\Gamma_0) =
\omega_0 \Bigg[ 1 + {\nu_{\rm x} \over 2} \bigg[1 - \bigg(
{\Gamma_0 \over \gamma_{\rm x}} \bigg)^2 \, \bigg] \Bigg]\,.
\label{eq:hf2}
\end{equation}
For the terminal point $A_{\rm f}$ of the anomalous dispersion
branch, which is characterized by $\kp^{\rm f} =
\{\mbox{Re}[\kp(\omega^{\rm f})],\mbox{Im}[\kp(\omega^{\rm
f})]\}$, one has $\mbox{Im}[\kp(\omega^{\rm f})] = 0$ and
$\mbox{Re}[\kappa] = 0$, and Eq.\,(\ref{eq:hf2}) yields
\begin{equation}
\mbox{Re}[\kp^{(2)}(\omega^{\rm f})] = k_0 \Bigg[ 1 + {\nu_{\rm x} \over 2} \bigg[1 - \bigg(
{\Gamma_0 \over \gamma_{\rm x}} \bigg)^2 \, \bigg] \Bigg]\,.
\label{eq:hf3}
\end{equation}

\begin{figure}[t]
\includegraphics[width=0.50\textwidth]{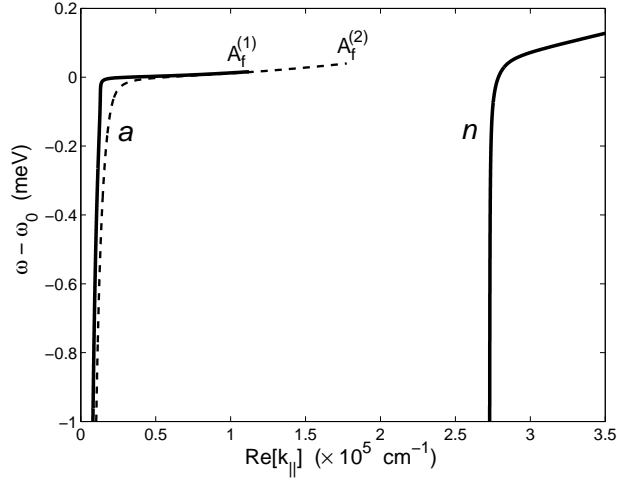}\\
\caption{The forced-harmonic dispersion branches of QW polaritons,
$\mbox{Re}[\kp] = \mbox{Re}[\kp(\omega)]$. The solid line $n$
refers to the normal QW polariton dispersion branch calculated
with Eq.\,(\ref{eq:dispersion}) for $\gamma_{\rm x} = 0$. The
solid (dashed) line $a$ is the anomalous, damping-induced
dispersion branch of QW polaritons evaluated for $\gamma_{\rm x} =
50\,\mu$eV ($\gamma_{\rm x} = 60\,\mu$eV). The in-plane
translational mass of QW excitons is given by $M_{\rm x} =
0.3\,m_0$, where $m_0$ is the free electron mass.}
\end{figure}

Similar to the case of the quasi-particle solutions, the
transition between the strong and weak coupling regimes of the QW
exciton -- photon interaction is attributed to the intersection of
the two dispersion curves, $k_{\|}^{(1)}(\omega)$ and
$k_{\|}^{(2)}(\omega)$, in 3D space
$\{\omega,\mbox{Im}(k_{\|}),\mbox{Re}(k_{\|})\}$ (see Fig.\,6).
Thus the topologically new dispersion branches ${\tilde
\kp}^{(1)}(\omega)$ and ${\tilde \kp}^{(2)}(\omega)$ arise from
the old ones, $k_{\|}^{(1)}(\omega)$ and $k_{\|}^{(2)}(\omega)$,
with $\gamma_{\rm x}$ increasing above $\tilde{\gamma}_{\rm
x}^{\rm tr} = \tilde{\gamma}_{\rm c}^{(2)}$. The transition point
is given by
\begin{eqnarray}
\tilde{\gamma}_{\rm x}^{\rm tr} &=& \tilde{\gamma}_{\rm c}^{(2)} \
= \ \frac{ 3 \sqrt{3}}{2}\,\left( \frac{\omega_0^2 R_{\rm QW}^2
\varepsilon_{\rm b}^2}{c^4 \hbar^2 M_{\rm x}}\right)^{1/3} = \,
\sqrt[3]{4} \, \nu_{\rm x}^{1/3} \gamma_{\rm c}^{(2)}\,,
\label{eq:hf4a}\\
\omega^{\rm tr} &=& \omega_0 \left[1 + {\omega_0 \varepsilon_{\rm
b} \over 2 c^2 M_{\rm x}} - { 1 \over \sqrt{3} \sqrt[3]{32} }\, {
\tilde{\gamma}_{\rm c}^{(2)} \over \omega_0 } \right]\,.
\label{eq:hf4b}
\end{eqnarray}

\begin{figure*}[t]
\resizebox{\hsize}{!}{
\begin{tabular}{c c c}
\includegraphics*[width=0.30\textwidth]{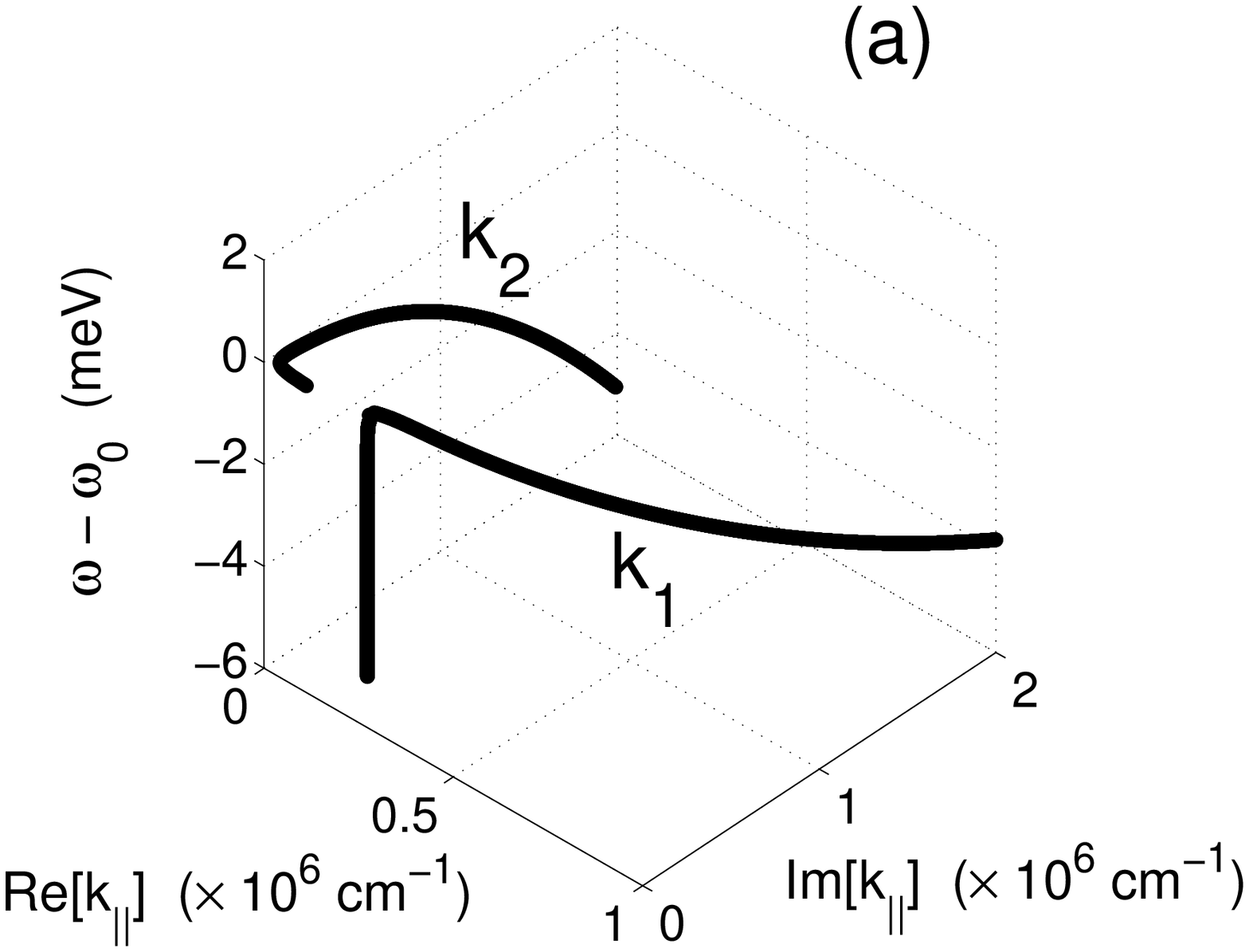}&
\includegraphics*[width=0.30\textwidth]{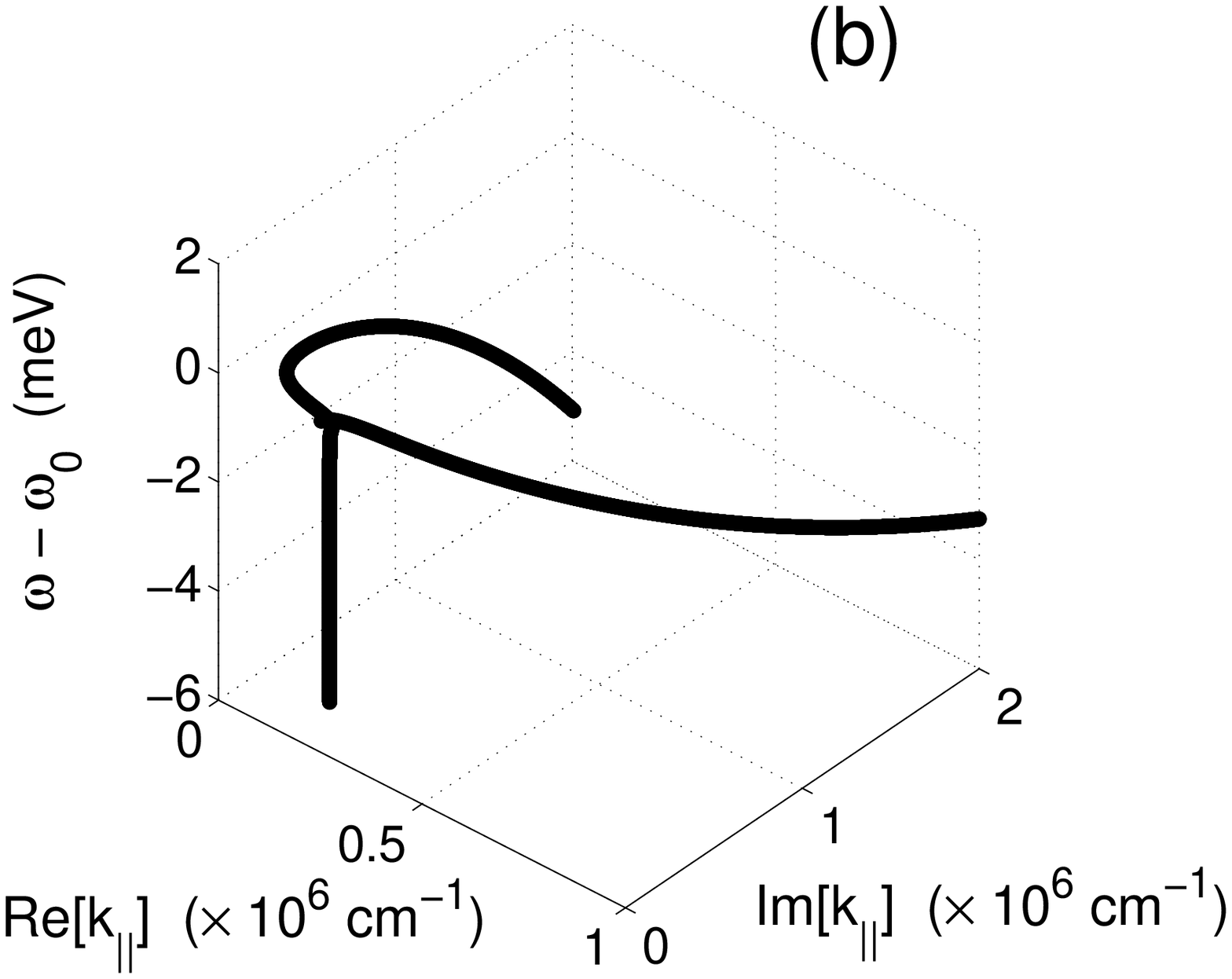}&
\includegraphics*[width=0.30\textwidth]{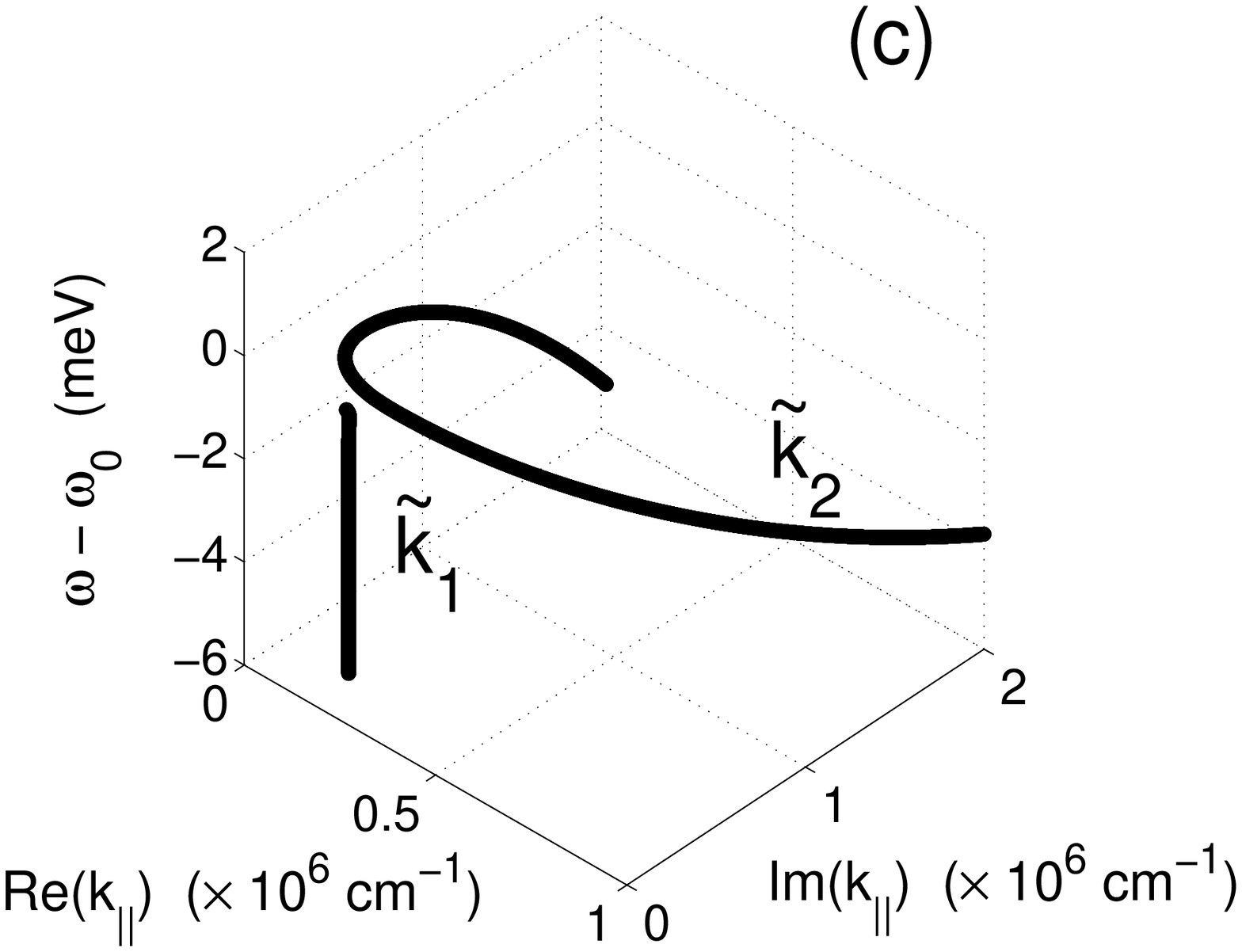}\\
\end{tabular} }
\caption{The $\gamma_{\rm x}$-induced transition between the
strong and weak coupling limits for forced-harmonic QW polaritons.
For $R_{\rm QW} = 0.025\,\mbox{eV}^2\AA$ and $M_{\rm x} =
0.3\,m_0$, the critical value of the damping rate is given by
$\tilde{\gamma}_{\rm x}^{\rm tr} = \tilde{\gamma}_{\rm c}^{(2)} =
119.8\,\mu$eV, according to Eq.\,(\ref{eq:hf4a}). (a) The strong
coupling regime, $\gamma_{\rm x} < \tilde{\gamma}_{\rm c}^{(2)}$,
(b) the transition point, $\gamma_{\rm x} = \tilde{\gamma}_{\rm
x}^{\rm tr} = \tilde{\gamma}_{\rm c}^{(2)}$, and (c) the weak
coupling regime, $\gamma_{\rm x} > \tilde{\gamma}_{\rm c}^{(2)}$.}
\end{figure*}

In contrast with the quasi-particle solution, for the
forced-harmonic solution the transition point is very sensitive to
the in-plane translational mass $M_{\rm x}$ of QW excitons:
$\tilde{\gamma}_{\rm x}^{\rm tr} \propto M_{\rm x}^{-1/3}$,
according to Eq.\,(\ref{eq:hf4a}). In particular, when $M_{\rm x}
\rightarrow \infty$ and therefore the spatial dispersion due to
excitons is removed, one has $\tilde{\gamma}_{\rm x}^{\rm tr} =
0$. In this case, the integrated absorption associated with the
exciton-like branch, ${\tilde \kp^{(2)}} = {\tilde
\kp^{(2)}}(\omega)$, is constant independent of the damping rate
$\gamma_{\rm x}$: $\int \mbox{Im}[{\tilde \kp^{(2)}}(\omega)] d
\omega = const \propto R_{\rm QW}$. This sum rule for the
absorption coefficient is known for bulk [\onlinecite{Loudon}] and
QW [\onlinecite{Andreanie}] excitons. Note that the large
difference between the critical damping rates $\gamma_{\rm x}^{\rm
tr} = \gamma_{\rm c}^{(2)}$ and $\tilde{\gamma}_{\rm x}^{\rm tr} =
\tilde{\gamma}_{\rm c}^{(2)}$ ($\gamma_{\rm x}^{\rm tr} \gg
\tilde{\gamma}_{\rm x}^{\rm tr}$, e.g., for our numerical
evaluations one has $\gamma_{\rm x}^{\rm tr} \simeq 2.4$\,meV
against $\tilde{\gamma}_{\rm x}^{\rm tr} \simeq 0.12$\,meV), which
refer to the quasi-particle (photoluminescence experiment) and
forced-harmonic (optical reflectivity/transmissivity experiment)
solutions, respectively, is well-known for bulk polaritons
\cite{Tait}.

\subsection{Optical brightness of damping-induced quantum well polaritons}

One can naturally question to what extent the $\gamma_{\rm
x}$-induced QW polariton dispersion branch is observable, i.e.,
``physical''. Some aspects of this question are discussed in
Sec.\,VI. Here, we analyze the photon component
$\varphi^{\gamma}_{\rm 2D}$ along the normal and anomalous QW
polariton dispersion branches. The photon component is a measure
of the brightness of polaritons.

In the initial Hamiltonian,
Eqs.\,(\ref{eq:hamiltoniana1})-(\ref{eq:H-parts}), quasi-2D
excitons couple with bulk photons. However, the confined QW
polariton modes deal with the quasi-2D light field: In this case
the QW excitons are dressed by the evanescent electromagnetic
field which in turn is trapped and guided by the QW exciton
states. The area-density of the electromagnetic energy $W^{\rm
2D}_{\rm phot}$, associated with the evanescent light field, is
given by
\begin{equation}
W_{\rm phot}^{\rm 2D} = \int_{-\infty}^{+\infty}\!\big( W_{\rm
E}^{\rm 3D} + W_{\rm H}^{\rm 3D} \big)\,dz =
\frac{1}{4\pi}\frac{|E(0)|^{2}}{\rm{Re}[\kappa]}\,,
\label{eq:2denergy}
\end{equation}
so that the dimensionless amplitude of the quasi-2D light field is
defined as
\begin{equation}
e^{\rm 2D}_{{\textbf{\textit k}}_{\|}} = \left( \frac{2}{L
\rm{Re}[\kappa]} \right)^{1/2} \sum_{k_{z}} \bigg( { \omega_0
\over \omega_{{\textbf{\textit k}}_{\|}}^{\gamma} } \bigg)^{1/2}
\! \big(\alpha_{{\textbf{\textit k}}_{\|},k_z} +
\alpha_{\textbf{\textit{-k}}_{\|},-k_z}^{\dag})\,. \label{eq:e0}
\end{equation}
The amplitude $e^{\rm 2D}_{{\textbf{\textit k}}_{\|}}$ can also be
interpreted as a combination of the creation and annihilation
operators of the quasi-2D photons. In a similar way, the
dimensionless amplitude of the quasi-2D excitonic polarization is
given by the operator $x^{\rm 2D}_{{\textbf{\textit k}}_{\|}} =
i(b_{{\textbf{\textit k}}_{\|}} - b^{\dag}_{-{\textbf{\textit
k}}_{\|}})$.

\begin{figure}[t!]
\includegraphics*[width=0.50\textwidth]{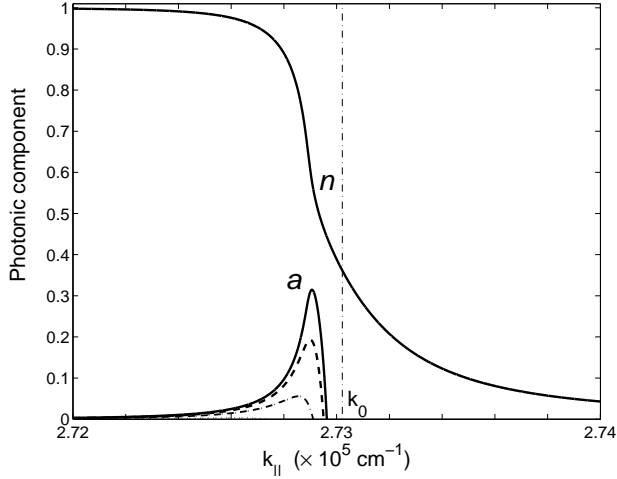}
\caption{The photon component $\varphi^{\gamma}_{\rm 2D} =
\varphi^{\gamma}_{\rm 2D}(k_{\|})$ of the normal and anomalous,
damping-induced branches for $\gamma_{\rm x} = 2.2$\,meV (solid
lines), and of the anomalous branch for $\gamma_{\rm x} =
2.0$\,meV (dashed line), $1.6$\,meV (dashed-dotted line), and
$1.0$\,meV (dotted line). The position of the photon cone, $k_{\|}
= k(\omega)$, is indicated by the vertical dash-dotted line.}
\end{figure}

Equations (\ref{eq:hamiltoniana1})-(\ref{eq:H-parts}) yield the
following relationship between $e^{\rm 2D}_{{\textbf{\textit
k}}_{\|}}$ and $x^{\rm 2D}_{{\textbf{\textit k}}_{\|}}$:
\begin{equation}
\big[ \omega^2 - (\omega^{\rm x}_{{\textbf{\textit k}}_{\|}})^2
\big] x^{\rm 2D}_{{\textbf{\textit k}}_{\|}} = \omega_0 \big(
R_{\rm QW} \mbox{Re}[\kappa] \big)^{1/2} e^{\rm
2D}_{{\textbf{\textit k}}_{\|}}\,. \label{eq:xe}
\end{equation}
The photon and exciton component of quantum well polaritons are
given by $\varphi^{\gamma}_{\rm 2D}(k_{\|}) = |\tilde{e}^{\rm
2D}_{{\textbf{\textit k}}_{\|}}|^2/(|\tilde{e}^{\rm
2D}_{{\textbf{\textit k}}_{\|}}|^2 + |\tilde{x}^{\rm
2D}_{{\textbf{\textit k}}_{\|}}|^2)$ and $\varphi^{\rm x}_{\rm
2D}(k_{\|}) = 1 - \varphi^{\gamma}_{\rm 2D} = |\tilde{x}^{\rm
2D}_{{\textbf{\textit k}}_{\|}}|^2/(|\tilde{e}^{\rm
2D}_{{\textbf{\textit k}}_{\|}}|^2 + |\tilde{x}^{\rm
2D}_{{\textbf{\textit k}}_{\|}}|^2)$, respectively. Here,
$\tilde{e}^{\rm 2D}_{{\textbf{\textit k}}_{\|}}$ and
$\tilde{x}^{\rm 2D}_{{\textbf{\textit k}}_{\|}}$ are the
phase-synchronous components of the electric and polarization
fields which contribute to the total stored energy of the quasi-2D
system. Thus, by using Eq.\,(\ref{eq:xe}), one derives:
\begin{equation}
\varphi^{\gamma}_{\rm 2D}(k_{\|}) = \frac{ \big(
\mbox{Re}[\omega^2 - (\omega^{\rm x}_{{\textbf{\textit
k}}_{\|}})^2] \big)^2 }{ \omega_0^2 R_{\rm QW} \mbox{Re}[\kappa] +
\big( \mbox{Re}[\omega^2 - (\omega^{\rm x}_{{\textbf{\textit
k}}_{\|}})^2] \big)^2 }\,. \label{eq:photcomp}
\end{equation}
For $\gamma_{\rm x}=0$, Eq.\,(\ref{eq:photcomp}) reduces to the
known expression for the photon component along the normal QW
polariton dispersion branch
[\onlinecite{Orrit,Ivanov97,Ivanov98}].

The photon component $\varphi^{\gamma}_{\rm 2D}$ along the
anomalous ($a$) and normal ($n$) QW polariton dispersion branches,
calculated with Eq.\,(\ref{eq:photcomp}) for the quasi-particle
solution of the dispersion Eq.\,(\ref{eq:dispersion}), is plotted
in Fig.\,7 for various damping rates $\gamma_{\rm x}
> \gamma_{\rm c}^{(1)}$. It is clearly seen that with increasing
$\gamma_{\rm x}$ the damping-induced QW polariton states become
bright, i.e., optically-active, only in close proximity of $k_{\|}
= k_0$. At the crossover point $k_{\|} = k_{\|}^{\rm tr}$ of the
dispersion branches, which occurs for $\gamma_{\rm x} =
\gamma_{\rm x}^{\rm tr} = \gamma_{\rm c}^{(2)}$ [see
Eqs.\,(\ref{eq:trans1})], the brightness of the anomalous and
normal QW polariton states becomes equal to each other. A similar
behaviour of the photon component $\varphi^{\gamma}_{\rm 2D}$ of
the QW polariton states takes place for the forced-harmonic
solution $k_{\|} = k_{\|}(\omega)$ of Eq.\,(\ref{eq:dispersion}).

\section{Strong-weak coupling transition for the
radiative states}

The damping-induced transition between the strong and weak
coupling regimes of the QW-exciton -- photon interaction can also
be traced for the radiative states. The changes of the radiative
width $\Gamma_{\rm T} = \Gamma_{\rm T}(\kp)$ and Lamb shift
$\Delta_{\rm T} = \Delta_{\rm T}(\kp)$ with increasing
$\gamma_{\rm x}$ are shown in Figs.\,8(a)-(b). Here, the radiative
corrections are defined as $\Gamma_{\rm T} = -2\mbox{Im}[\omega] -
\gamma_{\rm x}$ and $\Delta_{\rm T} = \mbox{Re}[\omega] -
\omega_0$, where $\omega$ is the solution of
Eq.\,(\ref{eq:dispersion}) that satisfies the following
conditions: $\mbox{Re}[\kappa] \equiv \mbox{Re}\big[\sqrt{\kp^{2}
- k^{2}(\omega)}\big] \leq 0$ and $\mbox{Im}[\omega] \leq 0$. The
solid lines in Figs.\,8(a)-(b) (see also the solid line in Fig.\,1
and the upper solid line in Fig.\,2) refers to the completely
coherent case, i.e., to the strong coupling limit with
$\gamma_{\rm x} = 0$, when the low-energy exciton state with
$k_{\|} \leq k_{\|}^{\rm (B)}$ is dressed by outgoing bulk photons
and interpreted as a radiative QW polariton. In this case, the
point $B$ (see Figs.\,1, 2, and 8) is the terminal point of the
dispersion of the radiative states.

For an arbitrary small $\gamma_{\rm x} > 0$ the dispersion of
radiative polaritons persists beyond the point $B$, where the
dispersion splits into two sub-branches, as a $\gamma_{\rm
x}$-induced tail of the radiative states at $k_{\|} \geq
k_{\|}^{\rm (B)}$. In this case, the upper dispersion sub-branch
beyond the bifurcation point $B$ becomes unphysical, i.e.,
completely disappears, while the damping-induced radiative tail is
associated with the lower dispersion sub-branch [see
Figs.\,8(a)-(b)]. The optical width $\Gamma_{\rm T}$ of the
$\gamma_{\rm x}$-induced radiative states at $k_{\|} \geq
k_{\|}^{\rm (B)}$ is approximated as
\begin{equation}
\Gamma_{\rm T}\big(\kp > \kp^{\rm (B)}\big) = \Gamma_0 \, \bigg(
\frac{\gamma_{\rm x}}{2\omega_0} \bigg) \bigg(
\frac{k_0^2}{\kp^{2}- k_{0}^{2}} \bigg)^{3/2} \,, \label{eq:tail}
\end{equation}
Approximation with Eq.\,(\ref{eq:tail}) is valid provided that
$(\hbar c/\sqrt{\eb}) \sqrt{\kp^{2}- k_{0}^{2}} \gg \gamma_{\rm
x}$. Equation (\ref{eq:tail}) clearly shows the damping-induced
nature of the radiative tail: $\Gamma_{\rm T}\big(\kp > \kp^{\rm
(B)}\big)$ is proportional to the incoherent damping rate
$\gamma_{\rm x}$ and therefore vanishes when $\gamma_{\rm x}
\rightarrow 0$. A similar $\gamma_{\rm x}$-induced tail of the
radiative states has also been numerically found for
quasi-one-dimensional plasmon-polaritons [\onlinecite{Citrin2}].

\begin{figure}[t!]
\includegraphics[width=0.5\textwidth]{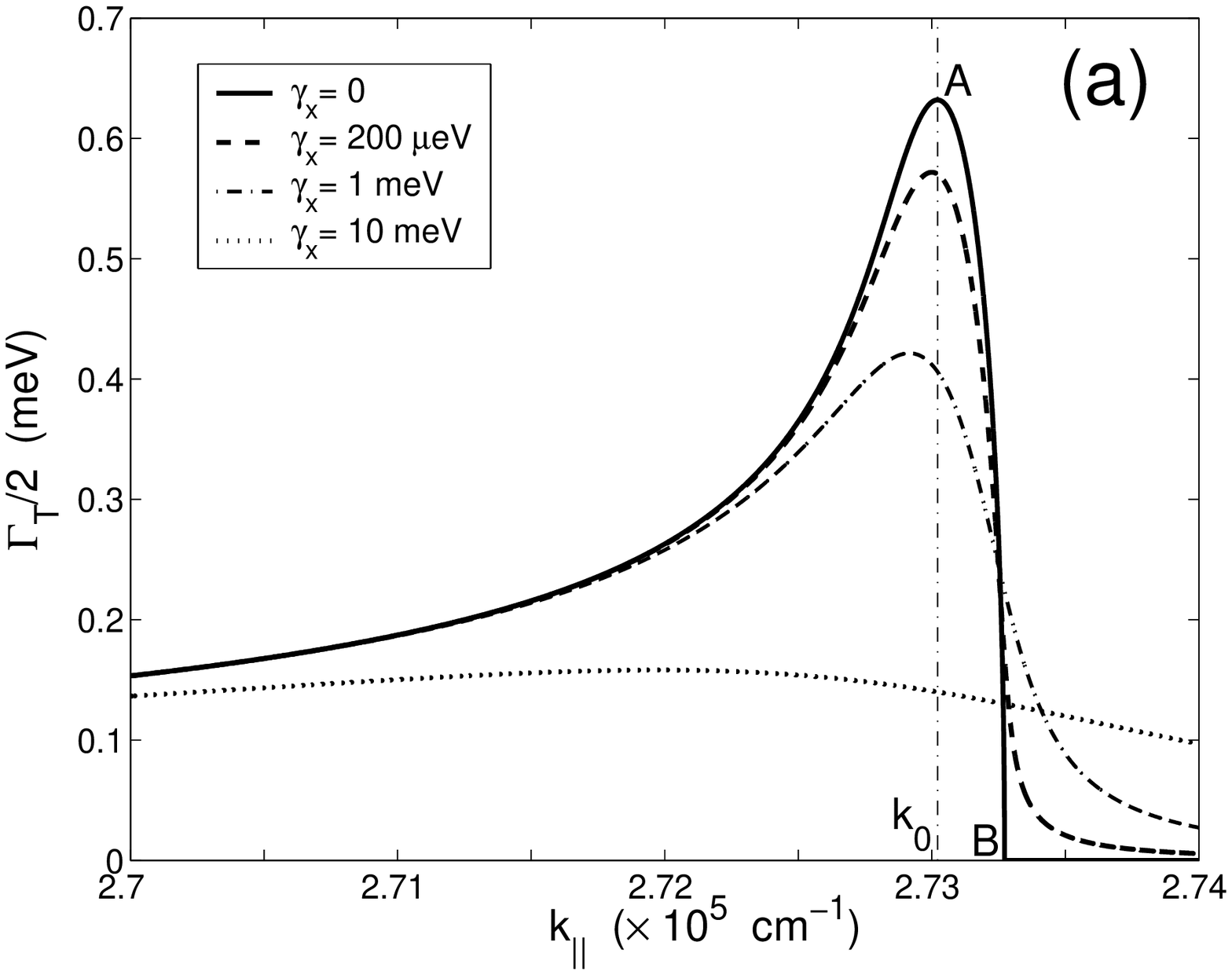}
\includegraphics[width=0.5\textwidth]{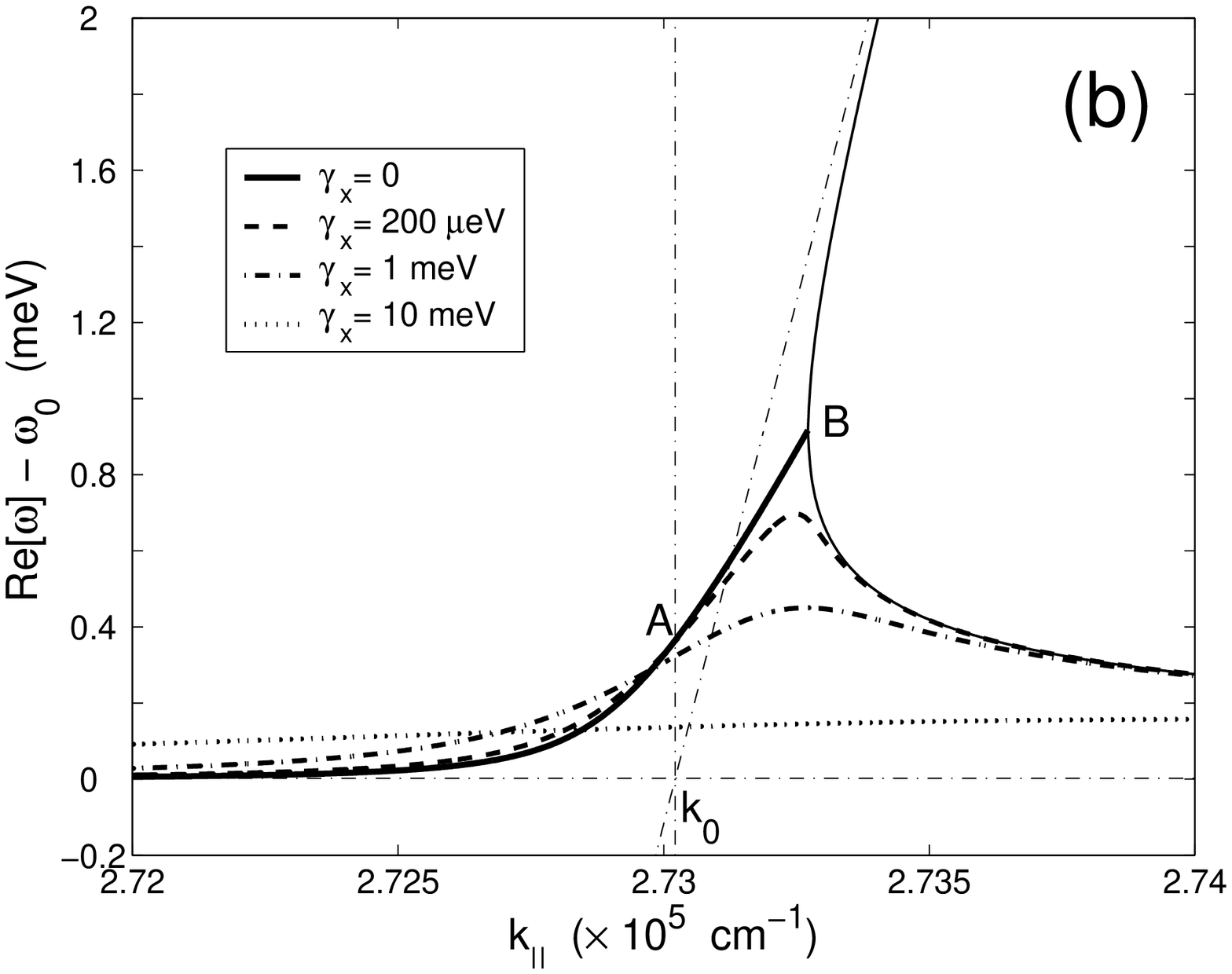}
\caption{
The dispersion of the $Y$-polarized radiative states (radiative
polaritons) for $\gamma_{\rm x} = 0$, 0.2\,meV, 1\,meV and
10\,meV: (a) the radiative half-width $\Gamma/2 = \Gamma_{\rm T}/2
= - \mbox{Im}[\omega(\kp)] - \gamma_{\rm x}/2$ and (b) the
radiative (Lamb) shift $\Delta = \Delta_{\rm T} =
\mbox{Re}[\omega(\kp)] - \omega_0$. For the QW parameters see the
caption of Fig.\,1.}
\end{figure}

As shown in Figs.\,8(a)-(b), in the vicinity of $\kp = k_0$ the
radiative corrections $\Gamma_{\rm T}$ and $\Delta_{\rm T}$
effectively decreases with increasing $\gamma_{\rm x}$. At the
same time, there is no damping-induced anomalous branch for the
radiative states of QW excitons. Therefore, in this case the
transition between the strong and weak coupling regimes of
exciton-photon interaction can only be approximately quantified in
terms of the $\gamma_{\rm x}$-induced qualitative changes of the
shape of the radiative corrections, $\Gamma_{\rm T} = \Gamma_{\rm
T}(\kp)$ and $\Delta_{\rm T} = \Delta_{\rm T}(\kp)$, at $\kp
\simeq k_0$ [see, e.g., the solid against dotted lines in
Figs.\,8(a)-(b)]. The drastic, qualitative changes occur when the
damping rate $\gamma_{\rm x}$ becomes comparable with the maximum
radiative corrections, $\Gamma^{\rm A}_{\rm T} \propto \big(
\Gamma_0^2 \omega_0 \big)^{1/3}$ given by Eq.\,(\ref{eq:A}) and
$\Delta^{\rm B}_{\rm T} \propto \big( \Gamma_0^2 \omega_0
\big)^{1/3}$ given by Eq.\,(\ref{eq:B}). In order to attribute the
transition to the critical damping $\gamma^{\rm tr}_{\rm x} =
\gamma_{\rm x}^{(2)} \propto \big( \Gamma_0^2 \omega_0
\big)^{1/3}$ defined by Eq.\,(\ref{eq:trans1}) for confined QW
polaritons, we choose the following criterion for the transition
point:
\begin{equation}
\gamma_{\rm x}^{\rm tr} = \gamma_{\rm c}^{(2)} =
\frac{3}{\sqrt[3]{4}}\,\Gamma_{\rm T}^{\rm A} \simeq
1.89\,\Gamma_{\rm T}^{\rm A} \,. \label{eq:tail2}
\end{equation}
In this case, the transition occurs synchronously for both
conjugated states, QW polaritons and radiative modes. As discussed
in Sec.\,VI, the transition can be visualized in the
angle-resolved photoluminescence signal from the radiative states
of QW excitons.

\section{DISCUSSION}

The anomalous, damping-induced dispersion branch of QW polaritons
can be visualized by using near-field or micrometric imaging
spectroscopy techniques \cite{Hess,Wu,Pulizzi} which allow to
detect the evanescent light field associated with the QW polariton
states. In Figs.~9 and 10 the evanescent field $E = E(z) = E(0)
e^{-\mbox{\small Re}[\kappa] |z|} \cos\big( \mbox{Im}[\kappa]
|z|\big)$ is plotted for the quasi-particle solution of the
dispersion Eq.\,(\ref{eq:dispersion}), for $\gamma_{\rm c}^{(1)} <
\gamma_{\rm x} < \gamma_{\rm c}^{(2)}$ and $\gamma_{\rm x} \simeq
\gamma_{\rm x}^{\rm tr} = \gamma_{\rm c}^{(2)}$, respectively. For
$\gamma_{\rm c}^{(1)} < \gamma_{\rm x} \ll \gamma_{\rm c}^{(2)}$,
the evanescent field associated with QW polaritons of the
anomalous dispersion branch spatially oscillates with the period
$2 \pi / \mbox{Im}[\kappa]$ which is much less than the
characteristic decay length of the envelope field,
$1/\mbox{Re}[\kappa]$ (see the solid line in Fig.\,9). A similar
behaviour of the evanescent field is well-known in physics of
surface acoustic waves \cite{Farnell}. In spite of spatial
oscillations of the evanescent field $E = E(z)$, there is,
however, no energy flux in the normal, $z$-direction: The
$\gamma_{\rm x}$-induced QW polariton states remain split off from
the radiative modes. For the strong-weak coupling transition
point, when $\gamma_{\rm x} = \gamma_{\rm x}^{\rm tr}$ and $\kp =
\kp^{\rm tr}$ [see Eq.\,(\ref{eq:trans1})], the evanescent fields
associated with normal and anomalous QW polaritons become
undistinguishable, as clearly illustrated in Fig.\,10.

\begin{figure}[t!]
\includegraphics[width=0.50\textwidth]{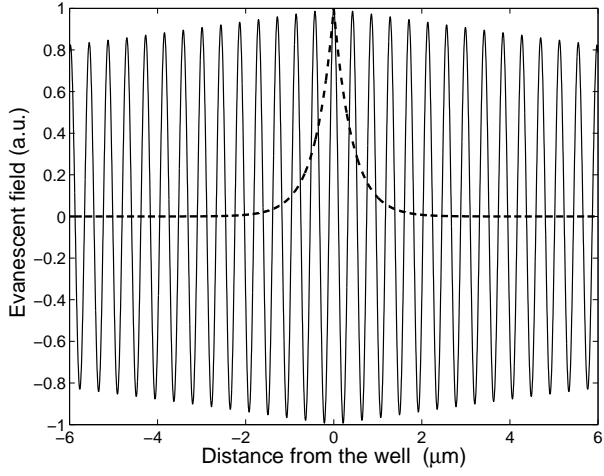}
\caption{The evanescent light field profile $E = E(z)$ associated
with normal (dashed line) and $\gamma_{\rm x}$-induced (solid
line) QW polaritons. The damping rate corresponds to $\gamma_{\rm
x} = 1$\,meV, while $\gamma^{(1)}_{\rm c} \simeq 0.046$\,meV and
$\gamma_{\rm x}^{\rm tr} = \gamma^{(2)}_{\rm c} \simeq 2.4$\,meV.}
\label{fig:profile1}
\end{figure}

\begin{figure}[t!]
\includegraphics[width=0.50\textwidth]{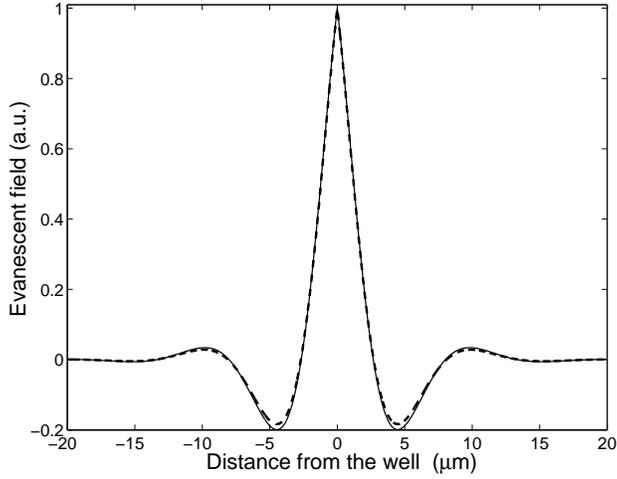}
\caption{The evanescent light field profile $E = E(z)$ associated
with normal (dashed line) and $\gamma_{\rm x}$-induced (solid
line) QW polaritons for the strong-weak coupling transition point:
$\gamma_{\rm x} = \gamma_{\rm x}^{\rm tr}$ and $\kp = \kp^{\rm
tr}$.} \label{fig:profile2}
\end{figure}

The transition between the strong and weak coupling regimes can
also be seen in photoluminescence of the QW radiative states at
grazing angles $\phi$ along the structure ($\phi \leqslant
2^{\circ}-5^{\circ}$ for GaAs QWs). The spectral function of the
PL signal is
\begin{equation}
S_{\rm PL} = S_{\rm PL}(\omega,\phi) = {1 \over \pi} \, \frac{
\varphi^{\rm x}(\kp) \mbox{Im}[\omega(\kp)]}{\big[
\mbox{Im}[\omega(\kp)] \big]^2 + \big[ \omega -
\mbox{Re}[\omega(\kp)] \big]^2 } \ ,
 \label{eq:spectral1}
\end{equation}
where $\omega(\kp)$ is the relevant complex solution of
Eq.\,(\ref{eq:dispersion}) and $\varphi^{\rm x}(\kp) =
\big|\mbox{Res}[G_{\kp}(\omega);\omega(\kp)]\big|$ is the exciton
component of the radiative states, both evaluated at $\kp =
\big[(\omega \sqrt{\eb})/(\hbar c) \big]\!\cos \phi$ for real
frequency $\omega$. Here, the exciton Green function, as a
solution of Eq.\,(\ref{eq:Dyson-equation}), is given by
\begin{equation}
G_{\textbf{\textit{k}}_{\parallel}}(\omega) = \frac{2
\omega_{\textbf{\textit{k}}_{\parallel}}^{\rm x}}{\omega^2 -
(\omega_{\textbf{\textit{k}}_{\parallel}}^{\rm x})^2  + (\omega^2
\eb R_{\rm QW})/(c^2 \hbar^2 \kappa) } \ .
 \label{eq:spectral2}
\end{equation}
The photoluminescence signal at grazing angles is proportional to
the spectral function, $I_{\rm PL} \propto S_{\rm
PL}(\omega,\phi)$, provided that the radiative states in the
vicinity of $\kp = k_0$ are equally populated.

\begin{center}
\begin{figure}[t!]
\includegraphics[width=0.70\textwidth]{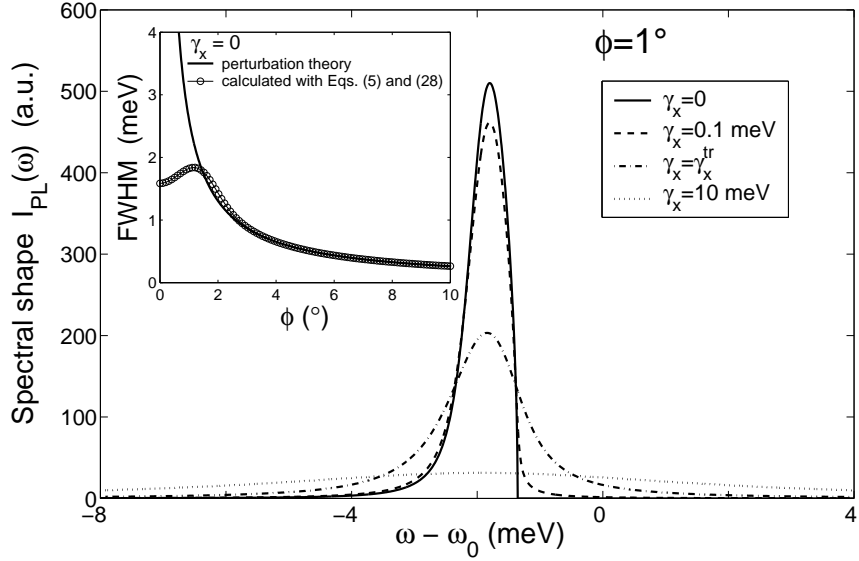}
\caption{The spectral shape of photoluminescence from the
radiative states of QW excitons, $I_{\rm PL} \propto S_{\rm
PL}(\omega,\phi)$, calculated with Eq.\,(\ref{eq:spectral1}) for
the grazing angle $\phi = 1^{\circ}$ along a GaAs singe quantum
well of 25\,nm thickness: $\gamma_{\rm x} = 0$ (solid line),
0.1\,meV (dashed line), $\gamma_{\rm x}^{\rm tr} = 2.4$\,meV
(dash-dotted line), and 10\,meV (dotted line). Inset: The
radiative width $\Gamma_{\rm T}$ against angle $\phi$, evaluated
with perturbation theory, by using Eq.\,(\ref{eq:gamma-per})
(dashed line), and calculated with Eqs.\,(\ref{eq:dispersion}) and
(\ref{eq:spectral1}) (solid line).}
\end{figure}
\end{center}

The spectral function $S_{\rm PL} = S_{\rm PL}(\omega)$ is plotted
in Fig.\,11 for the grazing angle $\phi = 1^{\circ}$ and various
rates $\gamma_{\rm x}$ of incoherent scattering. For $\gamma_{\rm
x} = 0$, the spectral function has a well-developed asymmetric
shape: $S_{\rm PL}(\omega) = 0$ for $\omega \geqslant \omega_0 +
\Delta_{\rm T}^{\rm (B)}$, where the Lamb shift $\Delta_{\rm
T}^{\rm (B)}$ is given by Eq.\,(\ref{eq:B}) (see also Figs.\,2 and
8), $S_{\rm PL}(\omega)$ rises very sharply with decreasing
$\omega$ right below $\omega_0 + \Delta_{\rm T}^{\rm (B)}$, and
has a Lorentzian-like red-side tail at $\omega < \omega_0$ (see
the solid line in Fig.\,11). With increasing $\gamma_{\rm x}$ the
spectral function $S_{\rm PL}(\omega)$ becomes more broad and
symmetric, and finally becomes Lorentzian-shaped at $\gamma_{\rm
x} \simeq \gamma_{\rm x}^{\rm tr}$ (see the dotted line in
Fig.\,11). The red shift of the maximum of $S_{\rm PL}(\omega)$
with increasing $\gamma_{\rm x}$ is consistent with the dependence
of the radiative corrections, $\Gamma_{\rm T}$ and $\Delta_{\rm
T}$, on $\gamma_{\rm x}$ (see Fig.\,8). In order to complete the
description, in the inset of Fig.\,11 we plot the the radiative
width at grazing angles, $\Gamma_{\rm T} = \Gamma_{\rm T}(\phi)$,
calculated for $\gamma_{\rm x} = 0$ by using standard perturbation
theory (dashed line) and evaluated with
Eqs.\,(\ref{eq:spectral1})-(\ref{eq:spectral2}) as the full width at
half maximum (FWHM) of the spectral function $S_{\rm PL} = S_{\rm
PL}(\omega,\phi)$ (solid line). In the former case the PL signal
from the radiative states diverges with $\phi \rightarrow 0$, as
$I_{\rm PL}(\phi) \propto \Gamma_{\rm T}(\phi) =
\Gamma_0/\sin\phi$, while the latter description yields removal of
the divergence. The described above $\gamma_{\rm x}$-induced
change of the spectral shape of the photoluminescence signal from
the radiative states can be seen at lager grazing angles, if,
e.g., ZnSe, CdTe, or GaN single QWs with a stronger oscillator
strength of exciton-photon interaction are used.

\begin{figure}[t!]
\includegraphics[width=0.50\textwidth]{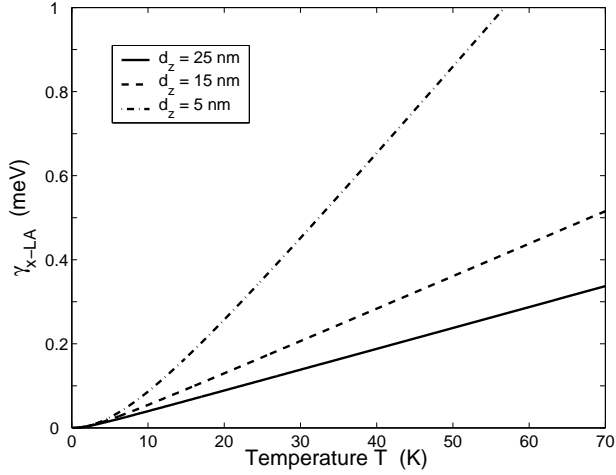}
\caption{The width $\gamma_{\rm x-LA} = \gamma_{\rm x-LA}(T)$
evaluated for GaAs single QWs of the thickness $d_z = 25$\,nm
(solid line), 15\,nm (dashed line), and 5\,nm (dash-dotted line).
Numerical calculations refer to the deformation potential of
exciton -- LA phonon interaction $D_{\rm DP} = 16.1$\,eV and the
in-plane translational mass $M_{\rm x} = 0.4\,m_0$ ($m_0$ is the
free electron mass).}
\end{figure}

The main mechanisms of incoherent scattering of QW excitons at low
temperatures are the deformation potential (DP) interaction of the
particles with bulk longitudinal (LA) phonons and exciton-exciton
interaction, so that $\gamma_{\rm x} = \gamma_{\rm x-LA} +
\gamma_{\rm x-x}$. In order to get an insight of how strong is the
first scattering channel, in Fig.\,12 we plot $\gamma_{\rm x-LA}$
against temperature $T$ for three values of the thickness of a
GaAs single quantum well, $d_z = 25$\,nm, 15\,nm, and 5\,nm. The
dependence $\gamma_{\rm x-LA} = \gamma_{\rm x-LA}(T,d_z)$, which
is calculated by using a method developed in
Ref.\,[\onlinecite{Stenius}], reflects the relaxation of the
momentum conservation in the $z$-direction for the QW exciton --
bulk LA phonon scattering process. For a low-density, classical
gas of QW excitons, when the bath temperature $T$ is much higher
than the degeneracy temperature $T_0$, the efficiency of QW
exciton -- QW exciton scattering is evaluated as
\begin{equation}
\gamma_{\rm x-x} = {\pi \over 4}\left(\frac{M_{\rm x}}{\mu_{\rm
x}}\right)^2 k_{\rm B} T_0 \ ,
 \label{eq:x-x}
\end{equation}
where $\mu_{\rm x}$ is the reduced mass of a QW exciton, $k_{\rm
B} T_0 = (\pi/2)(\hbar^2/M_{\rm x}) n_{\rm 2d}$, $n_{\rm 2d}$ is
the density of QW excitons, and the spin degeneracy factor of the
exciton states is $g=4$. According to Eq.\,(\ref{eq:x-x}),
$\gamma_{\rm x-x}$ is proportional to the concentration $n_{\rm
2d}$ of QW excitons, but, as a signature of the
quasi-two-dimensionality of the system, the QW exciton -- QW
exciton scattering rate is independent of the scattering length
and temperature $T$ (the latter is valid only for $T \gg T_0$).
For $n_{\rm 2d} = 10^9$\,cm$^{-2}$ and $M_{\rm x} = 0.4\,m_0$, the
degeneracy temperature $T_0 \simeq 35$\,mK, and
Eq.\,(\ref{eq:x-x}) yields $\gamma_{\rm x-x} \simeq 0.12$\,meV.
Figure~12 and estimates with Eq.\,(\ref{eq:x-x}) clearly show that
the majority of optical experiments
\cite{Deveaud1,Gurioli,Eccleston,Martinez,Vinattieri,Wu,Szczytko,Deveaud2}
with GaAs QWs are undertaken under conditions when the incoherent
scattering rate $\gamma_{\rm x}$ can easily achieve the values of
$\gamma_{\rm x}^{\rm tr}$ or ${\tilde \gamma}_{\rm x}^{\rm tr}$.

Another important question is to what extent the presented model
and results are robust against inhomogeneous broadening which is
practically inevitable in QW structures. To evaluate the influence
of the inhomogeneous broadening, we have examined our model in the
paradigm of a ``mean-field'' QW disorder, developed in
Ref.\,[\onlinecite{Andreanie}]: the results are valid provided
that the radiative corrections, $\Gamma_{\rm T}$ and $\Delta_{\rm
T}$, are much larger than the inhomogeneous broadening width,
$\delta_{\rm inhom}$. Because the radiative corrections relevant
to the strong-weak coupling transition refer to the in-plane
wavevector $\kp \simeq k_0$, they are large enough, $\Gamma_{\rm
T} \sim \Delta_{\rm T} \sim \big( \Gamma_0^2 \omega_0 \big)^{1/3}
\sim 1$\,meV, to meet the condition $\Gamma_{\rm T}, \Delta_{\rm
T} \gg \delta_{\rm inhom}$ for nowadays high-quality GaAs single
QWs.

\section{Conclusions}

In this paper we have studied how the $s$-polarized QW polariton
states, both confined and radiative, evolve in high-quality
(GaAs/AlGaAs) quantum wells with changing incoherent homogeneous
broadening $\gamma_{\rm x}$. The transition between the strong and
weak coupling regimes of the resonant QW exciton -- bulk photon
interaction is found and quantified. The following conclusions
summarize our results.

(i) In contrast with perturbation theory, there is no divergence
of the radiative width $\Gamma = \Gamma_{\rm T}(k_{\|})$ for
$k_{\|} \rightarrow k_0$. Furthermore, the radiative states of
$Y$-mode QW polaritons persist even beyond $k_0$, the crossover
point of the dispersions of bulk photons and QW excitons. The
regularization of the radiative corrections at $k_{\|} = k_0$
scales by the energy parameter $\big( \Gamma_0^2 \omega_0
\big)^{1/3} \gg \Gamma_0$, so that for $\Gamma_0 = 45.5\,\mu$eV
used in our numerical evaluations the maximum values of the
radiative corrections at $k_{\|} \simeq k_0$ are given by
$\Gamma_{\rm T}^{\rm max}/2 \simeq 0.63$\,meV and $\Delta_{\rm
T}^{\rm max} \simeq 0.92$\,meV.

(ii) For confined QW polaritons, a second, anomalous, $\gamma_{\rm
x}$-induced dispersion branch arises and develops with increasing
incoherent damping rate $\gamma_{\rm x}$. Such damping-induced
dispersion branches are known in plasma physics and in physics of
surface electromagnetic waves. In particular, for quasi-particle
solution $\omega = \omega(\kp)$ ($\kp$ is real) the critical value
of $\gamma_{\rm x}$ for the appearance of the second dispersion
branch $\omega = \omega_2(\kp)$ is $\gamma_{\rm x} = \gamma_{\rm
c}^{(1)} = \Gamma_0$. In this case, i.e., for $\gamma_{\rm x} \geq
\gamma_{\rm c}^{(1)}$, a finite sector of the dispersion branch
$\omega = \omega_2(k_{\|})$ penetrates into the physical part
(Re[$\kappa] \geq 0$ and Im[$\omega] \leq 0$) of 3D space
$\{k_{\|},\mbox{Im}[\omega],\mbox{Re}[\omega]\}$ from its
unphysical area.

(iii) The transition between the strong and weak coupling regimes
of the resonant QW exciton -- bulk photon interaction is
attributed to the incoherent scattering rate, $\gamma_{\rm x} =
\gamma_{\rm x}^{\rm tr}$ or $\tilde{\gamma}_{\rm x}^{\rm tr}$,
when intersection of the normal and damping-induced dispersion
branches of confined $s$-polarized QW polaritons occurs either in
$\{k_{\|},\mbox{Im}[\omega],\mbox{Re}[\omega]\}$ coordinate space
($\kp$ is real, the quasi-particle solution) or in
$\{\omega,\mbox{Im}[\kp],\mbox{Re}[\kp]\}$ coordinate space
($\omega$ is real, the forced-harmonic solution), respectively.
For the quasi-particle solution $\omega = \omega(\kp)$, the
transition damping rate is $\gamma_{\rm x}^{\rm tr} = \gamma_{\rm
c}^{(2)} \simeq 1.64\big( \Gamma_0^2 \omega_0 \big)^{1/3}$, and
the transition point is characterized by
Eqs.\,(\ref{eq:trans1})-(\ref{eq:trans2}). For the forced-harmonic
solution $\kp = \kp(\omega)$ ($\omega$ is real), one has
$\gamma^{\rm tr}_{\rm x} = $${\tilde \gamma}_{\rm c}^{(2)}$$\simeq
1.59 \big[ (\omega_0 \eb )/(M_{\rm x} c^2) \big]^{1/3} \big(
\Gamma_0^2 \omega_0 \big)^{1/3}$, and the transition point is
specified by Eq.\,(\ref{eq:hf4b}). Thus for a GaAs single quantum
well with the intrinsic radiative lifetime of QW excitons
$\tau_{\rm R} = 14.5$\,ps ($\Gamma_0 = 45.5\,\mu$eV) the
transition rates (widths) are given by $\gamma_{\rm x}^{\rm tr}
\simeq 2.39$\,meV and ${\tilde \gamma}_{\rm x}^{\rm tr} \simeq
120\,\mu$eV, respectively.

(iv) For confined QW polaritons, the evanescent light field
associated with normal-branch and $\gamma_{\rm x}$-induced-branch
polaritons as well as the transition between the strong and weak
coupling regimes can be visualized, e.g., by using near-field
optical spectroscopy, for both, quasi-particle and forced-harmonic
cases. For the radiative states of QW excitons, i.e., for
radiative QW polaritons, the transition (the quasi-particle case)
can be seen, e.g., in photoluminescence at grazing angles $\phi$,
as a qualitative change of the PL spectrum $I_{\rm PL} = I_{\rm
PL}(\omega,\phi)$ at a given $\phi$.
%and of the PL spatial
%diagram ${\tilde I}_{\rm PL} = {\tilde I}_{\rm PL}(\varphi) = \int
%I_{\rm PL}(\omega,\varphi) d\omega$.

\section{ACKNOWLEDGMENTS}

We appreciate valuable discussions with K. Cho, S. Elliott, L.
Mouchliadis, N.~I. Nikolaev, and I. Smolyarenko. Support of this
work by the EU RTN Project HPRN-CT-2002-00298 ``Photon-mediated
phenomena in semiconductor nanostructures'' is gratefully
acknowledged.


\end{document}